\documentclass[useAMS,usegraphicx,usenatbib]{mn2e}

\title[SN~2004aw]{SN~2004aw: Confirming Diversity of Type Ic Supernovae\thanks{Based on observations at ESO\,--Paranal, Prog. 074.D-0161(A)}}
\author[Taubenberger et al.]{S. Taubenberger$^{1}$,
A. Pastorello$^{1}$, P. A. Mazzali$^{1,2,3,4}$, S. Valenti$^{5,6}$, G. Pignata$^{7}$,
\newauthor D. N. Sauer$^{2}$, A. Arbey$^{8,1}$, O. B\"{a}rnbantner$^{9}$, S. Benetti$^{10}$, A. Della Valle$^{10}$,
\newauthor J. Deng$^{11,3,4}$, N. Elias-Rosa$^{10,12}$, A. V. Filippenko$^{13}$, R. J. Foley$^{13}$, A. Goobar$^{14}$,
\newauthor R. Kotak$^{5,15}$, W. Li$^{13}$, P. Meikle$^{15}$, J. Mendez$^{16,17}$, F. Patat$^{5}$, E. Pian$^{2}$, 
\newauthor C. Ries$^{9}$, P. Ruiz-Lapuente$^{16,1}$, M. Salvo$^{18}$, V. Stanishev$^{14}$,
\newauthor M. Turatto$^{10}$ and W. Hillebrandt$^{1}$\\
$^{1}$Max-Planck-Institut f\"{u}r Astrophysik, Karl-Schwarzschild-Str. 1, 85741 Garching bei M\"{u}nchen, Germany\\
$^{2}$INAF Osservatorio Astronomico di Trieste, Via Tiepolo 11, 34131 Trieste, Italy\\
$^{3}$Department of Astronomy, School of Science, University of Tokyo, Bunkyo-ku, Tokyo 113-0099, Japan\\
$^{4}$Research Center for the Early Universe, School of Science, University of Tokyo, Bunkyo-ku, Tokyo 113-0099, Japan\\
$^{5}$European Southern Observatory (ESO), Karl-Schwarzschild-Str. 2, 85748 Garching bei M\"{u}nchen, Germany\\
$^{6}$Physics Department, University of Ferrara, 44100 Ferrara, Italy\\
$^{7}$Departamento de Astronom\'ia y Astrof\'isica, Pontificia Universidad Cat\'olica de Chile, Casilla 306, Santiago 22, Chile\\
$^{8}$Universit\'e de Lyon 1, Centre de Recherche Astronomique de Lyon, 9 Avenue Charles Andr\'e, F-69230 Saint-Genis Laval, France\\
$^{9}$Universit\"{a}ts-Sternwarte M\"{u}nchen, Scheinerstr. 1, 81679 M\"{u}nchen, Germany\\
$^{10}$INAF Osservatorio Astronomico di Padova, Vicolo dell'Osservatorio 5, 35122 Padova, Italy\\
$^{11}$National Astronomical Observatories, Chinese Academy of Sciences, 20A Datun Road, Chaoyang District, Beijing 100012, China\\
$^{12}$Universidad de La Laguna, Av. Astrof\'isico Francisco S\'anchez s/n, E-38206 La Laguna, Tenerife, Spain\\
$^{13}$Department of Astronomy, University of California, Berkeley, CA 94720-3411, USA\\
$^{14}$Department of Physics, Stockholm University, AlbaNova University Center, SE-10691 Stockholm, Sweden\\
$^{15}$Astrophysics Group, Imperial College London, Blackett Laboratory, Prince Consort Road, London SW7 2AZ, UK\\
$^{16}$Department of Astronomy, University of Barcelona, Mart\'i i Franques 1, E-08028 Barcelona, Spain\\
$^{17}$Isaac Newton Group of Telescopes, 38700 Santa Cruz de La Palma, Islas Canarias, Spain\\
$^{18}$RSAA, Australian National University, Cotter Road, Weston Creek, ACT 2611, Australia}

\begin{document}

\date{Accepted 2006 July 5. Received 2006 July 4; in original form 2006 May 12}

\pagerange{\pageref{firstpage}--\pageref{lastpage}} \pubyear{2006}

\maketitle

\label{firstpage}

\begin{abstract}
Optical and near-infrared (near-IR) observations of the Type Ic 
supernova (SN Ic) 2004aw are presented, obtained from $-3$ to $+413$ 
days with respect to the $B$-band maximum. The photometric evolution 
is characterised by a comparatively slow post-maximum decline of the
light curves. The peaks in redder bands are significantly delayed
relative to the bluer bands, the $I$-band maximum occurring 8.4 days
later than that in $B$. With an absolute peak magnitude of $-18.02$ in
the $V$ band the SN can be considered fairly bright, but not
exceptional. This also holds for the $U$-through-$I$ bolometric light
curve, where SN~2004aw has a position intermediate between SNe~2002ap
and 1998bw. Spectroscopically SN~2004aw provides a link between a
normal Type Ic supernova like SN 1994I and the group of broad-lined
SNe~Ic. The spectral evolution is rather slow, with a spectrum at day
+64 being still predominantly photospheric. The shape of the nebular
[O\,{\sc i}] $\lambda\lambda6300,6364$ line indicates a highly
aspherical explosion. Helium cannot be unambiguously identified in the
spectra, even in the near-IR. Using an analytical description of
the light-curve peak we find that the total mass of the ejecta in
SN~2004aw is 3.5\,--\,8.0\,$M_\odot$, significantly larger than in
SN~1994I, although not as large as in SN~1998bw. The same model
suggests that about 0.3\,$M_\odot$ of $^{56}$Ni has been synthesised
in the explosion. No connection to a GRB can be firmly established.
\end{abstract}


\section{Introduction}
\label{Introduction}

Type Ic supernovae (SNe~Ic) were established as a distinct class in
the 1980s, and soon proposed to originate from the gravitational core
collapse of a massive star whose hydrogen and helium envelopes were
stripped through stellar winds or binary interaction \citep[for a
review on the different SN types see e.g.][]{Filippenko97}. However,
over the subsequent ten or more years only a few observational efforts
were made to learn more about the nature and properties of this
class of SNe. Exceptions were SN~1987M 
\citep{Filippenko90,Nomoto90,Jeffery91,Swartz93}, SN~1990U, SN~1990aa, and 
SN~1991A \citep{Filippenko92,Gomez94}, and especially SN~1994I
\citep[e.g.][]{Filippenko95,Richmond96}, which was studied in great
detail at various frequencies. For some other SNe~Ic, like SN~1983V
\citep{Clocchiatti97} and SN~1990B \citep{Clocchiatti01}, extensive data
sets were obtained, but their publication took some time and coincided
with the new boom in SN~Ic research, which set in with the discovery
of SN~1997ef \citep{IAUC6778,IAUC6786,IAUC6798,Iwamoto00,Mazzali00}
and SN~1998bw \citep{Galama98,Patat01}. These objects showed extremely
broad spectral features owing to an unusually high kinetic energy of
their ejecta, and therefore were labelled as ``hypernovae". Also, the
collection of SN~Ib/c spectra published by \citet{Matheson01} and
efforts to model spectra of SNe~Ib/c by \citet{Branch02} and 
\citet{Elmhamdi06} might be considered expressions of the new enthusiasm 
in this field.

In the following years, several more broad-lined SNe~Ic (hereafter
BL-SNe) were discovered. Unfortunately, a specific and generally
accepted definition of the term ``hypernova'' could not be
established; instead several different definitions were used in
parallel, creating considerable confusion. Probably the most commonly
adopted definition is that of \citet{Nomoto03,Nomoto05} and
\citet{Mazzali05a}, who refer to hypernovae as core-collapse SNe with
a total kinetic energy of the ejecta larger than $10^{52}$ erg, about
an order of magnitude larger than in ordinary SNe~Ic like SN~1994I. 
However, others use hypernova as a synonym for a jet-induced SN
connected to a gamma-ray burst (GRB) as predicted by the collapsar
model \citep{Paczynski98a,Paczynski98b,MacFadyen99}. In addition, a
classification based on observationally accessible quantities such as
the width of spectral features \citep{Mazzali02,Pandey03a} or the
luminosity \citep{Brown00} can be found in the literature. Because of
this ambiguity, we avoid the use of the term hypernova and rather
concentrate on the observable discrimination of SNe~Ic into
normal-velocity and broad-lined events on the basis of spectra
obtained near the time of optical maximum brightness.

Some of the SNe~Ic discovered so far could be linked to GRBs, while
for the majority there is apparently no association. Thus far, the
precise mechanism connecting GRBs with SNe is not fully understood. In
particular, it is not clear how we would infer the occurrence of a GRB
accompanying a SN explosion, if the former is directed away from the
observer, so that neither $\gamma$-rays nor $X$-rays are
detected. Given the relatively small number of well-observed SNe~Ic or
BL-SNe to date, every additional object with good photometric and
spectroscopic coverage may help to improve the picture of
stripped-envelope core-collapse SNe and their relation to GRBs. This
is the context of the observing campaign conducted on SN~2004aw.

SN~2004aw ($z$ = 0.0163) was discovered independently by
\citet{IAUC8310} on UT 2004 March 19.85 and 20.51, respectively. After
an initial classification as a SN~Ia \citep{IAUC8311,IAUC8312}, it was
reclassified as a SN~Ic \citep{IAUC8331}. The SN is located at
$\alpha$ = 11$^\rmn{h}$57$^\rmn{m}$50\fs25 and $\delta$ =
+25\degr15\arcmin55\farcs1, in a tidal tail of NGC 3997, a barred
spiral (SBb) galaxy and multiple system (LEDA\footnote{Lyon-Meudon
Extragalactic Database,\\ \hspace*{0.18cm}
\texttt{http://leda.univ-lyon1.fr/}}). \Citet{vandenBergh05} noted that
NGC 3997 is a merger system of two spiral galaxies showing tidally
deformed spiral arms.

\begin{figure}
   \centering
   \includegraphics{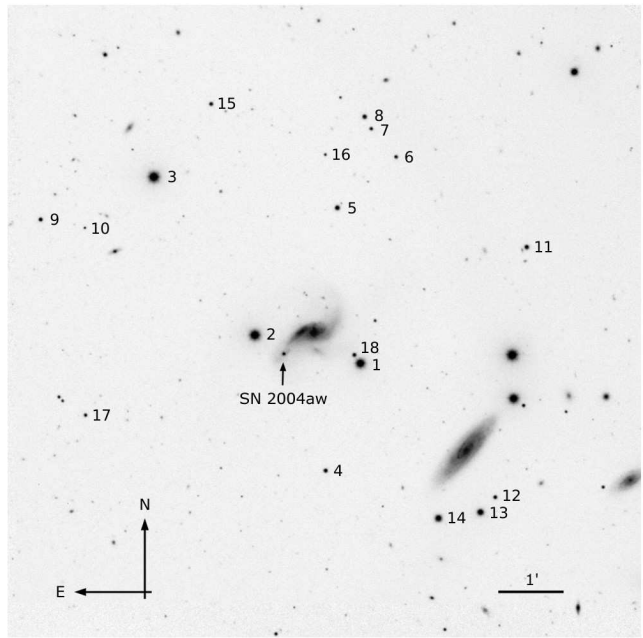}
   \caption{$J$~band image of the SN~2004aw field taken with the Calar
   Alto 3.5\,m Telescope + Omega 2000 on UT 2004 April 7. The local
   sequence stars are marked.}
   \label{fig:chart}
\end{figure}

After discovery and classification, the European Supernova
Collaboration\footnote{\texttt{http://www.mpa-garching.mpg.de/\,\~\,rtn/}}
and the Berkeley Supernova Group started follow-up observations in the
optical and infrared at various telescopes.  We present the entire set
of data obtained for SN~2004aw and discuss the techniques applied for
data reduction in Section~\ref{Data acquisition and reduction}. In
Section~\ref{Reddening and distance} we estimate the distance of
SN~2004aw and the extinction along the line of sight toward
it. Sections \ref{Light curves} and \ref{Spectral evolution} are
devoted to the analysis of the light curves and spectra,
respectively. Conclusions are drawn and their impact on the connection
between ordinary SNe~Ic, BL-SNe, and GRBs is discussed
in Section~\ref{Discussion}, followed by a short summary in
Section~\ref{Summary}. Synthetic light curves and spectra will be
presented in separate papers (Deng et al., in preparation; Baklanov et
al., in preparation).

\section{Data Acquisition and Reduction}
\label{Data acquisition and reduction}

\subsection{Photometry}
\label{Photometry}

Optical images of SN~2004aw were acquired from 3 days before to almost
1 year after maximum light (which occured on JD $2\,453\,088.4 \pm 0.5$
for the $B$ band), with dense sampling during the first two months.
Basic data reduction (bias subtraction, overscan correction, and flat-
fielding) was performed using standard routines in {\sc iraf}\footnote{{\sc iraf}
is distributed by the National Optical Astronomy Observatories, which
are operated by the Association of Universities for Research in
Astronomy, Inc, under contract to the National Science Foundation.}
\citep{IRAFphot,IRAFccdred}. A local sequence of stars in the SN field
(shown in Fig.~\ref{fig:chart}) was calibrated using observations of
several standard stars obtained on photometric nights.  Their
magnitudes, listed in Table~\ref{seq_mags}, were used to fix the
photometric zero-points for all nights and determine the SN relative
magnitudes. The stars labelled 1, 2, and 3 were saturated in almost
all images except those taken with the Katzman Automatic Imaging
Telescope (KAIT: Filippenko et al. 2001) at Lick Observatory, so they
were calibrated relative to the rest of the sequence from KAIT images
only.
\begin{table*}
\caption{Magnitudes of the local sequence stars in the field of
SN~2004aw (Fig.~\ref{fig:chart}).$^a$} 
\begin{footnotesize}
\begin{tabular}{ccccccccc}
\hline
ID & $U$ & $B$ & $V$ & $R$ & $I$ & $J$ & $H$ & $K'$\\
\hline
1 &                 & 13.67 $\pm$ 0.01& 13.01 $\pm$ 0.01& 12.64 $\pm$ 0.01& 12.40 $\pm$ 0.01&                 &                 &                 \\
2 &                 & 13.43 $\pm$ 0.01& 12.67 $\pm$ 0.01& 12.27 $\pm$ 0.01& 11.99 $\pm$ 0.01&                 &                 &                 \\
3 &                 & 13.62 $\pm$ 0.01& 12.71 $\pm$ 0.02& 12.21 $\pm$ 0.01& 11.79 $\pm$ 0.01&                 &                 &                 \\
4 & 20.77 $\pm$ 0.04& 19.44 $\pm$ 0.01& 18.09 $\pm$ 0.01& 17.20 $\pm$ 0.01& 16.38 $\pm$ 0.01& 15.27 $\pm$ 0.06& 14.68 $\pm$ 0.05& 14.53 $\pm$ 0.04\\
5 & 19.85 $\pm$ 0.04& 18.48 $\pm$ 0.01& 17.17 $\pm$ 0.02& 16.30 $\pm$ 0.03& 15.55 $\pm$ 0.02& 14.58 $\pm$ 0.08& 13.94 $\pm$ 0.07& 13.85 $\pm$ 0.09\\
6 &                 & 20.47 $\pm$ 0.03& 18.93 $\pm$ 0.02& 17.93 $\pm$ 0.02& 16.91 $\pm$ 0.01& 15.74 $\pm$ 0.04& 15.12 $\pm$ 0.08& 15.03 $\pm$ 0.06\\
7 & 17.61 $\pm$ 0.04& 17.77 $\pm$ 0.02& 17.27 $\pm$ 0.01& 16.95 $\pm$ 0.02& 16.64 $\pm$ 0.02& 16.23 $\pm$ 0.03& 15.93 $\pm$ 0.07& 16.00 $\pm$ 0.05\\
8 &                 & 19.93 $\pm$ 0.01& 18.41 $\pm$ 0.01& 17.34 $\pm$ 0.02& 16.19 $\pm$ 0.02& 14.95 $\pm$ 0.04& 14.34 $\pm$ 0.05& 14.23 $\pm$ 0.06\\
9 & 17.06 $\pm$ 0.05& 17.22 $\pm$ 0.03& 16.68 $\pm$ 0.02& 16.28 $\pm$ 0.03& 15.93 $\pm$ 0.03& 15.52 $\pm$ 0.03& 15.21 $\pm$ 0.03& 15.22 $\pm$ 0.04\\
10& 21.00 $\pm$ 0.06& 19.98 $\pm$ 0.06& 19.04 $\pm$ 0.02& 18.44 $\pm$ 0.02& 17.94 $\pm$ 0.05&                 &                 &                 \\
11& 19.24 $\pm$ 0.05& 18.11 $\pm$ 0.04& 16.95 $\pm$ 0.03& 16.28 $\pm$ 0.03& 15.75 $\pm$ 0.04& 14.81 $\pm$ 0.04& 14.22 $\pm$ 0.04& 14.15 $\pm$ 0.03\\
12& 18.04 $\pm$ 0.02& 17.82 $\pm$ 0.01& 17.08 $\pm$ 0.01& 16.70 $\pm$ 0.02& 16.30 $\pm$ 0.02& 15.64 $\pm$ 0.04& 15.32 $\pm$ 0.03& 15.24 $\pm$ 0.06\\
13& 16.38 $\pm$ 0.02& 15.66 $\pm$ 0.02& 14.76 $\pm$ 0.01& 14.30 $\pm$ 0.02& 13.88 $\pm$ 0.02& 13.16 $\pm$ 0.02& 12.80 $\pm$ 0.04& 12.77 $\pm$ 0.03\\
14& 18.91 $\pm$ 0.02& 17.63 $\pm$ 0.01& 16.13 $\pm$ 0.02& 15.16 $\pm$ 0.01& 14.18 $\pm$ 0.02&                 &                 &                 \\
15&                 &                 &                 &                 &                 & 15.48 $\pm$ 0.03& 14.90 $\pm$ 0.05& 14.76 $\pm$ 0.04\\
16&                 &                 &                 &                 &                 & 16.95 $\pm$ 0.04& 16.37 $\pm$ 0.08& 16.29 $\pm$ 0.11\\
17&                 &                 &                 &                 &                 & 16.09 $\pm$ 0.02& 15.48 $\pm$ 0.08& 15.39 $\pm$ 0.05\\
18&                 &                 &                 &                 &                 & 15.37 $\pm$ 0.04& 14.73 $\pm$ 0.05& 14.60 $\pm$ 0.05\\
\hline\\[-0.7ex]
\end{tabular}
$^a$The optical data for stars 4 to
14 were obtained on 9 photometric nights using CAFOS, DOLORES, and
PFIP, while stars 1, 2, and 3 were calibrated with KAIT images
relative to the rest of the sequence. In the near-IR, all nights were
used to calibrate the sequence.
\end{footnotesize}
\label{seq_mags}
\end{table*}

The instrumental SN magnitudes were determined with point-spread
function (PSF) fitting photometry using the software package ``{\sc snoopy},"
specifically designed for this purpose by F. Patat and implemented in
{\sc iraf} by E. Cappellaro. In general, no template galaxy subtraction was
done, because no suitable templates were available. For the KAIT
images, templates obtained on days +266 to +270 with the same
instrument and under variable seeing conditions were used to perform
galaxy subtraction. This technique is expected to provide superior
results given the non-isolated position of SN~2004aw within its host
galaxy and the mostly poor seeing of the KAIT observations. However,
the SN had not entirely faded away at that epoch, which caused a flux
over-subtraction. This was corrected by subtracting from the KAIT
templates the SN flux measured in the nearly simultaneous (day +258)
TNG observations, which have better quality and seeing than those
acquired with KAIT. Depending on epoch and filter, the corrections
derived in this way range from 0.02 to 0.10 mag.

We noted that in some of our observations the filter response deviates
from the standard Johnson/Cousins passbands \citep{Bessell90}. Thus,
besides a first-order colour-term correction derived from observations
of Landolt standard fields, we applied an additional correction based
on the prescription of \citet{Stritzinger02}, \citet{Pignata04b}, and
references therein. These ``$S$-corrections" were computed for the
epochs from 0 to +57 days in the $R$ and $I$ bands and from 0 to +36
days in the $B$ and $V$ bands. Only for the Wendelstein Telescope and
the William Herschel Telescope, for which we did not have the required
filter response curves, the $S$-corrections were omitted. The derived
corrections were mostly small (typically $\leq 0.03$ mag), but in some
cases reached values of up to $\sim$\,0.10 mag. The largest
corrections were applied to the $B$- and $I$-band magnitudes obtained
with CAFOS. In summary, the corrections were very modest in the $V$
and $R$ bands, they produced a small reduction of the scatter in the
$B$ band, and improved substantially the coherence of the $I$-band
magnitudes.

The final apparent magnitudes for SN~2004aw are reported in
Table~\ref{SN_mags}, together with their uncertainties, computed as
the sum in quadrature of the measurement errors of the instrumental SN
magnitudes and the error associated with the photometric zero-point of
the night.

\begin{table*}
\caption{Photometric observations of SN~2004aw.$^a$} 
\label{SN_mags}
\begin{footnotesize}
\begin{tabular}{lrrcclllll}
\hline
UT Date  & JD\ \ \ \ \ & Epoch$^b$ & \qquad $U$ & \qquad $B$ & \qquad $V$ & \qquad $R$ & \qquad $I$ & Telescope & \ Seeing$^c$ \\
         & $- 2\,453\,000$ & (days)\ \    &            &            &            &            &            &           & (arcsec) \\
\hline
04/03/21 &  85.8\ \ \ \ & $-2.6$\ \ \ \ &              & 18.12$\pm$0.04 & 17.47$\pm$0.04 & 17.19$\pm$0.04 & 16.97$\pm$0.05 & \ \ KAIT & \ \ 2.24\\
04/03/23 &  87.8\ \ \ \ & $-0.6$\ \ \ \ &              & 18.08$\pm$0.03 & 17.36$\pm$0.04 & 17.07$\pm$0.02 & 16.77$\pm$0.03 & \ \ KAIT & \ \ 2.64\\
04/03/24 &  88.8\ \ \ \ &  0.4\ \ \ \ &                & 18.06$\pm$0.12 & 17.32$\pm$0.11 & 16.97$\pm$0.06 & 16.78$\pm$0.11 & \ \ KAIT & \ \ 2.58\\
04/03/24$^d$ &  89.4\ \ \ \ &  1.0\ \ \ \ & 18.18$\pm$0.06 & 18.07$\pm$0.06 & 17.29$\pm$0.04 & 17.01$\pm$0.04 & 16.71$\pm$0.04 & \ \ TNG  & \ \ 1.38\\
04/03/26 &  90.9\ \ \ \ &  2.5\ \ \ \ &                &                & 17.31$\pm$0.03 & 16.93$\pm$0.06 & 16.63$\pm$0.04 & \ \ KAIT & \ \ 2.69\\
04/03/28 &  92.8\ \ \ \ &  4.4\ \ \ \ &                & 18.38$\pm$0.07 & 17.32$\pm$0.04 & 16.91$\pm$0.04 & 16.60$\pm$0.04 & \ \ KAIT & \ \ 3.90\\
04/03/28 &  93.1\ \ \ \ &  4.7\ \ \ \ & 18.38$\pm$0.11 & 18.24$\pm$0.06 & 17.31$\pm$0.03 & 16.87$\pm$0.03 & 16.52$\pm$0.02 & \ \ SSO  & \ \ 2.62\\
04/03/29 &  93.8\ \ \ \ &  5.4\ \ \ \ &                & 18.39$\pm$0.04 & 17.35$\pm$0.04 & 16.92$\pm$0.04 & 16.54$\pm$0.04 & \ \ KAIT & \ \ 2.44\\
04/03/29 &  94.1\ \ \ \ &  5.7\ \ \ \ & 18.55$\pm$0.05 & 18.35$\pm$0.05 & 17.37$\pm$0.06 & 16.91$\pm$0.04 & 16.54$\pm$0.07 & \ \ SSO  & \ \ 2.20\\
04/03/29 &  94.5\ \ \ \ &  6.1\ \ \ \ &                & 18.42$\pm$0.11 & 17.44$\pm$0.06 & 16.99$\pm$0.06 & 16.54$\pm$0.06 & \ \ Ekar & \ \ 3.16\\
04/03/30 &  94.8\ \ \ \ &  6.4\ \ \ \ &                & 18.43$\pm$0.10 & 17.33$\pm$0.06 & 16.91$\pm$0.07 & 16.53$\pm$0.07 & \ \ KAIT & \ \ 2.48\\
04/03/30 &  95.5\ \ \ \ &  7.1\ \ \ \ &                & 18.52$\pm$0.10 & 17.38$\pm$0.06 & 16.88$\pm$0.04 & 16.43$\pm$0.05 & \ \ WD   & \ \ 1.57\\
04/03/30 &  95.5\ \ \ \ &  7.1\ \ \ \ &                &                & 17.43$\pm$0.16 & 
& 16.54$\pm$0.06 & \ \ NOT  & \ \ 0.75\\
04/03/31 &  95.8\ \ \ \ &  7.4\ \ \ \ &                & 18.53$\pm$0.08 & 17.43$\pm$0.06 & 16.92$\pm$0.03 & 16.54$\pm$0.04 & \ \ KAIT & \ \ 2.32\\
04/03/31 &  96.5\ \ \ \ &  8.1\ \ \ \ & 18.95$\pm$0.09 & 18.58$\pm$0.04 & 17.32$\pm$0.05 & 16.85$\pm$0.04 & 16.40$\pm$0.05 & \ \ NOT  & \ \ 0.96\\
04/04/01 &  96.8\ \ \ \ &  8.4\ \ \ \ &                & 18.56$\pm$0.12 & 17.44$\pm$0.05 & 16.95$\pm$0.04 & 16.56$\pm$0.05 & \ \ KAIT & \ \ 3.08\\
04/04/01 &  97.4\ \ \ \ &  9.0\ \ \ \ &                & 18.81$\pm$0.08 & 17.34$\pm$0.08 & 16.88$\pm$0.03 & 16.42$\pm$0.03 & \ \ WD   & \ \ 1.82\\
04/04/07 & 102.8\ \ \ \ & 14.4\ \ \ \ &                & 
& 17.69$\pm$0.06 & 17.04$\pm$0.04 & 16.58$\pm$0.04 & \ \ KAIT & \ \ 2.44\\
04/04/07 & 103.3\ \ \ \ & 14.9\ \ \ \ &                & 19.14$\pm$0.05 & 17.71$\pm$0.05 & 17.07$\pm$0.05 & 16.59$\pm$0.05 & \ \ Caha & \ \ 1.34\\
04/04/07$^d$ & 103.5\ \ \ \ & 15.1\ \ \ \ & 19.81$\pm$0.18 & 19.25$\pm$0.08 & 17.73$\pm$0.11 & 17.04$\pm$0.05 & 16.56$\pm$0.10 & \ \ TNG  & \ \ 1.93\\
04/04/09 & 104.8\ \ \ \ & 16.4\ \ \ \ &                & 19.25$\pm$0.07 & 17.82$\pm$0.03 & 17.10$\pm$0.05 & 16.65$\pm$0.03 & \ \ KAIT & \ \ 2.24\\
04/04/12 & 107.8\ \ \ \ & 19.4\ \ \ \ &                & 19.47$\pm$0.11 & 18.02$\pm$0.15 & 17.26$\pm$0.05 & 16.71$\pm$0.05 & \ \ KAIT & \ \ 2.74\\
04/04/12 & 108.3\ \ \ \ & 19.9\ \ \ \ &                & 19.49$\pm$0.06 & 17.97$\pm$0.03 & 17.23$\pm$0.03 & 16.71$\pm$0.03 & \ \ Caha & \ \ 1.47\\
04/04/14$^d$ & 109.6\ \ \ \ & 21.2\ \ \ \ &      & 19.54$\pm$0.11 & 18.01$\pm$0.04 & 17.27$\pm$0.04 & 16.74$\pm$0.04 & \ \ Caha & \ \ 1.91\\
04/04/14$^d$ & 110.4\ \ \ \ & 22.0\ \ \ \ &      & 19.60$\pm$0.03 & 18.07$\pm$0.03 & 17.33$\pm$0.04 & 16.77$\pm$0.05 & \ \ Caha & \ \ 1.14\\
04/04/16 & 111.8\ \ \ \ & 23.4\ \ \ \ &                & 19.67$\pm$0.12 & 18.15$\pm$0.05 & 17.41$\pm$0.04 & 16.88$\pm$0.03 & \ \ KAIT & \ \ 2.84\\
04/04/17 & 113.4\ \ \ \ & 25.0\ \ \ \ &                & 19.77$\pm$0.15 & 18.22$\pm$0.07 & 17.49$\pm$0.06 & 16.93$\pm$0.05 & \ \ Caha & \ \ 3.70\\
04/04/18 & 114.5\ \ \ \ & 26.1\ \ \ \ & 20.81$\pm$0.08 & 19.95$\pm$0.04 & 18.30$\pm$0.03 & 17.55$\pm$0.03 &                & \ \ NOT  & \ \ 0.86\\
04/04/20 & 116.4\ \ \ \ & 28.0\ \ \ \ & 20.91$\pm$0.06 & 19.96$\pm$0.06 & 18.35$\pm$0.03 & 17.61$\pm$0.03 & 17.00$\pm$0.03 & \ \ NOT  & \ \ 0.69\\
04/04/20 & 116.4\ \ \ \ & 28.0\ \ \ \ &                & 19.99$\pm$0.05 & 18.42$\pm$0.04 & 17.65$\pm$0.03 & 17.03$\pm$0.04 & \ \ Caha & \ \ 1.09\\
04/04/21 & 117.4\ \ \ \ & 29.0\ \ \ \ &                & 20.17$\pm$0.15 & 18.44$\pm$0.05 & 17.67$\pm$0.07 & 17.03$\pm$0.07 & \ \ Ekar & \ \ 2.30\\
04/04/21 & 117.5\ \ \ \ & 29.1\ \ \ \ &                & 20.02$\pm$0.05 & 18.46$\pm$0.04 & 17.68$\pm$0.04 & 17.05$\pm$0.05 & \ \ Caha & \ \ 1.57\\
04/04/23 & 118.6\ \ \ \ & 30.2\ \ \ \ &                & 20.04$\pm$0.09 & 18.47$\pm$0.07 & 17.71$\pm$0.07 & 17.08$\pm$0.04 & \ \ Caha & \ \ 2.92\\
04/04/23$^d$ & 119.4\ \ \ \ & 31.0\ \ \ \ &      & 20.09$\pm$0.09 & 18.52$\pm$0.06 & 17.74$\pm$0.04 & 17.12$\pm$0.05 & \ \ Caha & \ \ 1.84\\
04/04/24 & 119.8\ \ \ \ & 31.4\ \ \ \ &                & 20.05$\pm$0.11 & 18.57$\pm$0.07 & 17.78$\pm$0.05 & 17.15$\pm$0.05 & \ \ KAIT & \ \ 2.46\\
04/04/24$^d$ & 120.5\ \ \ \ & 32.1\ \ \ \ & 20.88$\pm$0.14 & 20.14$\pm$0.05 & 18.58$\pm$0.04 & 17.80$\pm$0.03 & 17.13$\pm$0.03 & \ \ Caha & \ \ 1.35\\
04/04/25$^d$ & 121.5\ \ \ \ & 33.1\ \ \ \ &      & 20.20$\pm$0.07 & 18.64$\pm$0.03 & 17.83$\pm$0.03 & 17.16$\pm$0.04 & \ \ Caha & \ \ 1.30\\
04/04/29 & 124.8\ \ \ \ & 36.4\ \ \ \ &                & 20.49$\pm$0.43 & 18.86$\pm$0.20 & 17.99$\pm$0.06 & 17.29$\pm$0.08 & \ \ KAIT & \ \ 3.18\\
04/05/04 & 129.8\ \ \ \ & 41.4\ \ \ \ &                & 20.37$\pm$0.41 & 18.75$\pm$0.11 & 18.05$\pm$0.08 & 17.33$\pm$0.06 & \ \ KAIT & \ \ 2.68\\
04/05/06$^d$ & 132.5\ \ \ \ & 44.1\ \ \ \ & 21.43$\pm$0.13 & 20.53$\pm$0.06 & 18.97$\pm$0.04 & 18.15$\pm$0.02 & 17.52$\pm$0.04 & \ \ TNG  & \ \ 1.05\\
04/05/09 & 134.8\ \ \ \ & 46.4\ \ \ \ &                & 20.47$\pm$0.19 & 19.01$\pm$0.09 & 18.25$\pm$0.08 & 17.53$\pm$0.06 & \ \ KAIT & \ \ 2.54\\
04/05/11$^d$ & 137.4\ \ \ \ & 49.0\ \ \ \ & 21.44$\pm$0.13 & 20.63$\pm$0.07 & 19.09$\pm$0.03 & 18.29$\pm$0.02 & 17.62$\pm$0.03 & \ \ WHT  & \ \ 1.38\\
04/05/13 & 138.7\ \ \ \ & 50.3\ \ \ \ &                & 20.51$\pm$0.17 & 19.05$\pm$0.08 & 18.31$\pm$0.06 & 17.59$\pm$0.06 & \ \ KAIT & \ \ 2.20\\
04/05/13 & 139.4\ \ \ \ & 51.0\ \ \ \ &                & 20.65$\pm$0.11 & 19.16$\pm$0.05 & 18.32$\pm$0.04 &                & \ \ Ekar & \ \ 2.06\\
04/05/20 & 145.7\ \ \ \ & 57.3\ \ \ \ &                & 
& 19.15$\pm$0.07 & 18.44$\pm$0.05 & 17.77$\pm$0.09 & \ \ KAIT & \ \ 2.08\\
04/05/27 & 152.7\ \ \ \ & 64.3\ \ \ \ &                & 
& 19.32$\pm$0.16 & 18.63$\pm$0.10 & 17.81$\pm$0.10 & \ \ KAIT & \ \ 2.18\\
04/06/03 & 159.7\ \ \ \ & 71.3\ \ \ \ &                & 
& 19.33$\pm$0.25 & 18.57$\pm$0.12 & 17.91$\pm$0.08 & \ \ KAIT & \ \ 2.68\\
04/12/06 & 346.7\ \ \ \ &258.3\ \ \ \ &   $>$ 23.3    & 23.45$\pm$0.34 & 22.70$\pm$0.37 & 21.26$\pm$0.23 & 20.84$\pm$0.12 & \ \ TNG  & \ \ 1.43\\
05/03/06 & 436.7\ \ \ \ &348.3\ \ \ \ &                &   $>$ 24.0    &                & 22.27$\pm$0.44 &                & \ \ VLT  & \ \ 1.76\\
\hline
\end{tabular}
\\[1.2ex]
$^a$ The magnitudes are $S$-corrected, but not corrected for
interstellar extinction. The KAIT images acquired 266 to 270 days past
maximum are not listed as they did not allow any measurement of the SN
magnitudes.\\
$^b$ Epoch with respect to the estimated $B$-band maximum JD
2\,453\,088.4 $\pm$ 0.5.\\
$^c$ Average seeing over all filter bands.\\
$^d$ Night used to calibrate a local sequence.\\[1.5ex]
KAIT = 0.76\,m Katzman Automated Imaging Telescope; pixel scale = 0.8"/px\\
TNG = 3.58\,m Telescopio Nazionale Galileo + Dolores; pixel scale = 0.275"/px\\
SSO = Siding Spring Observatory 2.3\,m Telescope + Imager; pixel scale = 0.60"/px\\
Ekar = Asiago 1.82\,m Telescope + AFOSC; pixel scale = 0.473"/px\\
WD = 0.8\,m Wendelstein Telescope + Monica; pixel scale = 0.50"/px\\
NOT = 2.5\,m Nordic Optical Telescope + ALFOSC; pixel scale = 0.188"/px\\
Caha = Calar Alto 2.2\,m Telescope + CAFOS SiTe; pixel scale = 0.53"/px\\
WHT = 4.2\,m William Herschel Telescope + PFIP; pixel scale = 0.24"/px\\
VLT = ESO 8.2\,m Very Large Telescope + FORS2; pixel scale = 0.252"/px
\end{footnotesize}
\end{table*}

IR photometry in the $JHK'$ bands was obtained using NICS mounted on
TNG and Omega 2000 mounted on the Calar Alto 3.5\,m Telescope. The
observations range from approximately 10 to 29 days after maximum
light.  While the basic reduction (sky determination and subtraction,
combination of dithered images) was always performed manually in order
to have better control over changes in the sky background, a dedicated
tool ({\sc snap}\footnote{Written by F. Mannucci,\\
\hspace*{0.18cm} \texttt{http://www.arcetri.astro.it/\,\~\,filippo/snap/}}) 
was used to correct cross-talking and field distortions in the NICS frames.

\begin{table*}
\caption{IR photometry of SN~2004aw.$^a$}
\label{IR_mags}
\begin{center}
\begin{footnotesize}
\begin{tabular}{lrrlllll}
\hline
UT Date  & JD\ \ \ \ \ & Epoch$^b$ & \qquad $J$ & \qquad $H$ & \qquad $K'$ & Telescope & \ Seeing$^c$ \\
         & $- 2\,453\,000$ & (days)\ \    &            &            &            &            & (arcsec) \\
\hline
04/04/03 &  98.6\ \ \ \ & 10.2\ \ \ \ & 15.98$\pm$0.06 & 15.78$\pm$0.06 & 15.65$\pm$0.08 & \ \ TNG  & \ \ 0.96\\
04/04/07 & 102.6\ \ \ \ & 14.2\ \ \ \ & 15.88$\pm$0.03 & 15.79$\pm$0.06 & 15.64$\pm$0.07 & \ \ Caha & \ \ 1.35\\
04/04/08 & 103.6\ \ \ \ & 15.2\ \ \ \ & 15.87$\pm$0.06 & 15.80$\pm$0.06 & 15.72$\pm$0.06 & \ \ Caha & \ \ 1.06\\
04/04/13 & 109.4\ \ \ \ & 21.0\ \ \ \ & 15.92$\pm$0.03 & 15.86$\pm$0.04 & 15.72$\pm$0.07 & \ \ Caha & \ \ 0.96\\
04/04/21 & 117.4\ \ \ \ & 29.0\ \ \ \ & 16.25$\pm$0.09 & 15.99$\pm$0.14 & 15.95$\pm$0.13 & \ \ TNG  & \ \ 0.96\\
\hline
\end{tabular}
\end{footnotesize}
\end{center} 
\begin{footnotesize}
$^a$ The magnitudes are not corrected for interstellar extinction.\\
$^b$ Epoch with respect to the estimated $B$-band maximum JD 2\,453\,088.4 $\pm$ 0.5.\\
$^c$ Average seeing over all filter bands.\\[1.1ex] 
TNG = Telescopio Nazionale Galileo + NICS; pixel scale = 0.25"/px\\ 
Caha = Calar Alto 3.5\,m Telescope + Omega 2000; pixel scale = 0.45"/px
\end{footnotesize}
\end{table*}

As in the case of optical photometry, the SN magnitudes were
determined with respect to a calibrated sequence of stars in the
field. These are the same as for optical wavelengths, complemented by
four additional red stars that are either not detected in most optical
bands or are close to another bright star. Their calibration was
performed with standard-field observations \citep{Hunt98} at both
telescopes, and with the help of the magnitudes reported in the
2MASS\footnote{\texttt{http://www.ipac.caltech.edu/2mass/index.html}}
catalogue. For the latter comparison, the $K$-band magnitudes were
transformed to the $K'$ band following \citet{Wainscoat92}.  The SN
measurements were performed using the PSF-fitting technique; no
template galaxy subtraction was done.  Instrumental magnitudes were
transformed to the standard $JHK'$ system \citep{Bessell88,
Wainscoat92} by adjusting the zero-points, but without applying any
colour correction. The latter is the most uncertain aspect of our
calibration, but all our attempts to determine the colour terms of the
two instruments with the help of standard fields or the 2MASS local
sequence yielded inconsistent values. Therefore, we have omitted colour
corrections, but account for this deficiency with an additional
contribution to the total photometric error. The derived $JHK'$
magnitudes and their uncertainties are shown in Table~\ref{IR_mags}.

\subsection{Spectroscopy}
\label{Spectroscopy}

The spectroscopic observations are reported in
Table~\ref{spectra}. Until SN~2004aw starts its transition to the
nebular phase at $\sim$\,2 months from maximum light, its evolution is
densely sampled at optical wavelengths. During the nebular phase, the
SN was recovered and three more spectra were taken at phases between
236 and 413 days after maximum. 

\begin{table*}
\caption{Spectroscopic observations of SN~2004aw.}
\begin{center}
\begin{footnotesize}
\begin{tabular}{lrrclclcl}
\hline
UT Date   & JD\ \ \ \ \  & Epoch$^a$ & Airmass & Telescope & Grism\,/\,Grating  & \quad Range        & Resolution$^b$ & Standards \\
          & $- 2\,453\,000$  & (days)\ \ &         &           &                    & \quad \ \ (\AA)    & (\AA)          &           \\[0.8ex]
\hline
04/03/24  &  89.4\ \ \ \ & 1.0\ \ \ \  &  1.21 & \ TNG-D   & LR-B\,+\,LR-R      &  3200 --  9200     & 12             & Feige56   \\
04/03/28  &  93.0\ \ \ \ & 4.6\ \ \ \  &  1.83 & \ SSO     & grt300\,+\,grt316  &  3650 --  9000     &  5             & Feige56, Cd329927\\
04/03/29  &  94.0\ \ \ \ & 5.6\ \ \ \  &  1.85 & \ SSO     & grt300\,+\,grt316  &  3650 --  9000     &  5             & Feige56, L745-46 \\
04/03/29  &  94.5\ \ \ \ & 6.1\ \ \ \  &  1.26 & \ Ekar    & gm\,4              &  3800 --  7700     & 24             &           \\
04/03/31  &  96.5\ \ \ \ & 8.1\ \ \ \  &  1.01 & \ NOT     & gm\,4              &  3500 --  8900     & 14             & Feige67   \\
04/04/07  & 103.5\ \ \ \ & 15.1\ \ \ \ &  1.01 & \ TNG-D   & LR-B\,+\,LR-R      &  3500 --  9400     & 14             & Hd93521   \\
04/04/13  & 109.5\ \ \ \ & 21.1\ \ \ \ &  1.06 & \ Caha    & b200\,+\,r200      &  3550 --  9200     & 10             & Feige66   \\
04/04/14  & 110.5\ \ \ \ & 22.1\ \ \ \ &  1.04 & \ Caha    & b200\,+\,r200      &  3500 --  9600     & 10             & Feige66   \\
04/04/18  & 114.5\ \ \ \ & 26.1\ \ \ \ &  1.21 & \ NOT     & gm\,4              &  3650 --  8900     & 13             &           \\
04/04/20  & 116.4\ \ \ \ & 28.0\ \ \ \ &  1.12 & \ Caha    & b200\,+\,r200      &  3500 --  9400     & 10             & Feige66   \\
04/04/21  & 116.8\ \ \ \ & 28.4\ \ \ \ &  1.01 & \ UKIRT   & gm\,$H\!K$         & $\!\!\!13700$ -- 24500 & 34         & BS4501 (F4V)\\
04/04/21  & 117.4\ \ \ \ & 29.0\ \ \ \ &  1.16 & \ TNG-N   & Amici prism        &  7500 -- 25000     & 60-200         & AS24-0 (A0)\\
04/04/21  & 117.4\ \ \ \ & 29.0\ \ \ \ &  1.26 & \ Ekar    & gm\,4              &  3800 --  7700     & 24             & Bd+332642 \\
04/04/22  & 118.4\ \ \ \ & 30.0\ \ \ \ &  1.26 & \ Ekar    & gm\,2              &  5200 --  9400     & 75             & Bd+332642 \\
04/04/27  & 123.5\ \ \ \ & 35.1\ \ \ \ &  1.26 & \ Caha    & b200               &  3800 --  8650     & 12             & Feige66   \\
04/05/06  & 132.4\ \ \ \ & 44.0\ \ \ \ &  1.05 & \ TNG-D   & LR-R               &  4950 --  9400     & 11             & Feige56   \\
04/05/12  & 137.8\ \ \ \ & 49.4\ \ \ \ &  1.21 & \ Lick    & gm600\,+\,grt300   &  5200 --  9600     & 5/11           & Feige34, HD84937\\ 
04/05/25  & 151.4\ \ \ \ & 63.0\ \ \ \ &  1.54 & \ Ekar    & gm\,2              &  5200 --  9200     & 75             &           \\
04/05/27  & 152.8\ \ \ \ & 64.4\ \ \ \ &  1.45 & \ Lick    & grt600             &  4000 --  6650     &  8             & HD84937   \\
04/11/14  & 324.1\ \ \ \ & 235.7\ \ \ \ & 1.50 & \ Keck    & gm400\,+\,grt400   &  3750 --  9300     &  6             & Feige34, Bd+174708\\
04/12/09  & 348.8\ \ \ \ & 260.4\ \ \ \ & 1.08 & \ TNG-D   & LR-B               &  3800 --  7950     & 15             & GD140     \\
05/05/11  & 501.8\ \ \ \ & 413.4\ \ \ \ & 1.04 & \ Keck    & gm400\,+\,grt400   &  5450 --  9200     &  6             & Feige34, HD84937\\
\hline\\[-3.0ex]
\end{tabular}
\end{footnotesize}
\end{center} 
\begin{footnotesize}
$^a$ Relative to $B$-band maximum (JD = 2\,453\,088.4).\\
$^b$ Full-width at half maximum (FWHM) of isolated, unblended
night-sky lines.\\[1.6ex]
TNG-D = 3.58\,m Telescopio Nazionale Galileo + Dolores\\
TNG-N = 3.58\,m Telescopio Nazionale Galileo + NICS\\
SSO = Siding Spring Observatory 2.3\,m Telescope + DBS\\
NOT = 2.5\,m Nordic Optical Telescope + ALFOSC\\
Caha = Calar Alto 2.2\,m Telescope + CAFOS SiTe\\
Ekar = Asiago 1.82\,m Telescope + AFOSC\\
UKIRT = 3.8\,m United Kingdom Infrared Telescope + UIST\\
Lick = Shane 3\,m Reflector + Kast Dual Spectrograph\\
Keck = Keck 10\,m Telescope + LRIS
\end{footnotesize}
\label{spectra}
\end{table*}

All optical frames were first debiased and flat-fielded before the
spectra were optimally extracted \citep{IRAFspec} using standard {\sc iraf}
routines. Wavelength calibration was accomplished with the help of 
arc-lamp exposures or, whenever this was not possible or unsatisfactory,
using the night-sky lines. The instrumental sensitivity functions
required for flux calibration were determined from observations of the
spectrophotometric standard stars reported in
Table~\ref{spectra}. Whenever no standard was observed, the
sensitivity curve obtained on a different night with the same
instrumental configuration was used. Telluric features were recognised
in the spectra of the spectrophotometric standard stars and removed
from the SN spectra. 

Most of our SN~2004aw and standard-star spectra were obtained at low 
airmass (see Table~\ref{spectra}) and with the slit along the parallactic 
angle \citep{Filippenko82}; surely, whenever the airmass exceeded 
$\sim$\,1.3 the parallactic angle was used.
Therefore, the shape of the continua should not be strongly 
affected by differential flux losses. Moreover, all spectra were 
checked against the photometry, and multiplication by a constant factor 
to correct for flux losses due to a misalignment of the slit or 
to clouds proved to be sufficient to achieve a good agreement. 
The two spectra taken on days +5.6 and +6.1 have been co-added to 
increase the signal-to-noise ratio (S/N).

Almost contemporaneously with the last epoch of IR photometry (i.e.,
29 days after maximum light in the $B$ band), two IR spectra were
obtained (see Table~\ref{spectra}). The one taken with the TNG/Amici
prism covers the full wavelength range from the $I$-through-$K$ bands,
but has extremely low resolution (FWHM of night-sky lines $\approx
60-200$~\AA). The other, taken at UKIRT, has rather poor S/N.

Both spectra were reduced following standard procedures. After
pairwise subtraction of dithered frames, the SN traces were optimally
extracted, scaled to the flux level of the spectrum with the highest
S/N (assuming that this suffered the least flux loss), and then
combined. Wavelength calibration was performed using arc-lamp
exposures (UIST) or a tabulated function relating pixel number to
wavelength (NICS). Removal of telluric features and a rough flux
calibration were accomplished with the help of stars of spectral type
A0 and F4V for NICS and UIST, respectively. In order to construct the
sensitivity curves containing both instrumental response and
atmospheric absorption, the A0 star was compared to Vega, and the F4V
spectrum to that of the Sun. The final flux calibration was done with
respect to the simultaneous SN photometry, and a constant correction
factor was sufficient to provide satisfactory agreement.  Finally, the
two IR spectra were averaged in their common wavelength range (the $H$
and $K$ bands).

\section{Distance and Extinction}
\label{Reddening and distance}

This section deals with two aspects that turn out to represent the
largest source of uncertainty in the absolute calibration of the
SN~2004aw data. Neither Cepheid distances nor other similarly precise
distance measurements are available for NGC 3997, the host galaxy of
SN~2004aw. Therefore, the redshift is the only indicator of its
distance. LEDA reports a recession velocity of 4769 km\,s$^{-1}$
inferred mainly from radio measurements, which increases to 4906
km\,s$^{-1}$ when a correction for the Local-Group infall onto
the Virgo cluster is applied.  The recession velocities of the other
members of the small galaxy cluster to which NGC 3997 belongs show
only little dispersion (approximately $\pm 250$ km\,s$^{-1}$ around an
average Local-Group infall corrected value of 4700 km\,s$^{-1}$). Thus,
assuming 500 km\,s$^{-1}$ as an upper limit for the peculiar motion
seems conservative, and we adopt this value for the uncertainty in
determining the distance from the Hubble law. Using $v_\rmn{rec}$ =
4906 km\,s$^{-1}$ and $H_0=72\,\rmn{km}\,\rmn{s}^{-1}\rmn{Mpc}^{-1}$,
this implies a distance modulus of $\mu=34.17\pm0.23$ mag.

The uncertainty associated with the amount of extinction must be
considered the dominant source of error in the calibration of the
SN~2004aw data. Based on our current knowledge from a very limited set
of well-observed nearby objects, SNe~Ic have very heterogeneous
observed properties. Hence, in contrast to SNe~Ia
\citep[e.g.][]{Phillips99} and to a certain degree also SNe~IIP
\citep{Hamuy03,Pastorello}, it is not possible to infer the amount of
reddening from light curves or colour curves of SNe~Ic, so that (apart
from spectral modelling) the only possibility of getting an estimate
of the extinction along the line of sight towards SN~2004aw is by
relating the equivalent width (EW) of the interstellar Na\,{\sc
i}\,D lines in the spectrum to the colour excess. This procedure
suffers from the entirely unknown composition of the dust, and in
particular the unconstrained gas-to-dust ratio of the extinguishing
material. In addition, the lines might be saturated, which is
impossible to be verified in low-resolution spectra. 

For SN~2004aw we find a moderately strong Na\,{\sc i}\,D line at the
redshift of the host galaxy, and in the few spectra with sufficiently
high S/N a weak line at its rest wavelength is also visible. The
equivalent widths were determined to be 2.17 $\pm$ 0.11~\AA\ and 0.65
$\pm$ 0.03~\AA, respectively (errors are only statistical). Using
\begin{equation}
E(B-V)\, =\, 0.16\ {\rm EW}\textrm{(Na\,{\sc i}\,D)}\,,
\end{equation}
a revised version of the relation presented by \citet{Turatto_proc},
this implies a reddening $E(B-V)$ = 0.35 $\pm$ 0.02 mag in NGC
3997, and a Galactic reddening of $E(B-V)$ = 0.10 $\pm$ 0.01 mag. The
value for the Galactic contribution is inconsistent with the colour
excess of $E(B-V)$ = 0.021 mag reported by \citet{Schlegel98} for the
direction towards NGC 3997. The discrepancy could either be attributed
to small-scale variations in the dust distribution of the Milky Way,
to scatter in the Turatto et al. relation, or to a measurement error
owing to the weakness of the line. Since we consider this last option 
to be the most likely, we adopt the Galactic reddening of Schlegel et
al., which yields a total reddening along the line of sight of 0.37
mag. With a value of 3.1 typically assumed for $R_V$ = $A_V/E(B-V)$
(for a more detailed discussion see,
e.g., \citealt{Cardelli89,O'Donnell94,Riess96,Phillips99,Elias-Rosa06})
we obtain $A_V$ = 1.15 mag and $A_B$ = 1.52 mag.  We conservatively
assume a total uncertainty in our estimate of the $B-V$ colour excess of
25\% or 0.10 mag.

\section{Light curves}
\label{Light curves}

\subsection{Optical light curves}
\label{Optical light curves}

\begin{figure}   
   \centering
   \includegraphics[width=84mm]{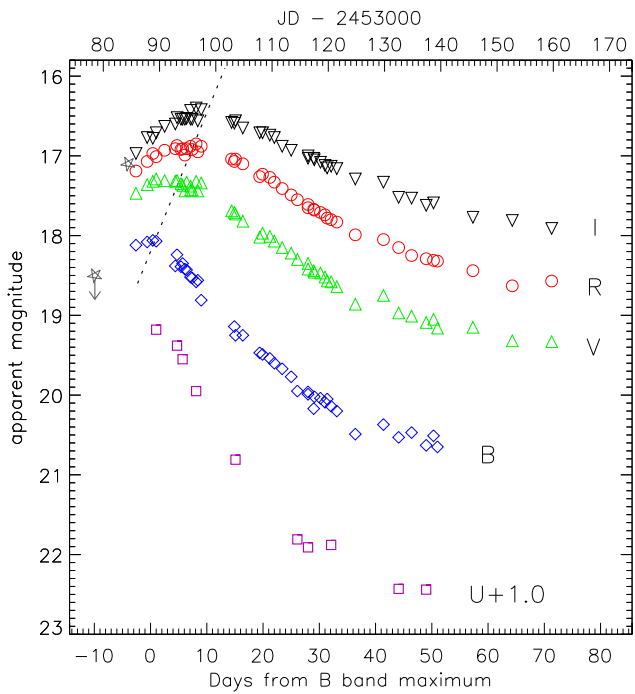}
   \caption{$U\!BV\!RI$ light curves of SN~2004aw from $-3$ to +71
days from $B$-band maximum (which is JD 2\,453\,088.4 $\pm$ 0.5). In
addition, two unfiltered measurements (open starred symbols) made by
amateur astronomers are shown: T. Boles reports a magnitude of $17.1$ for JD
2\,453\,084.4 (discovery) and K. Itagaki provides a detection limit of
$18.5$ for JD 2\,453\,078.5 \citep{IAUC8310}. The dotted line roughly
connects the maxima of the $B$-through-$I$ bands.}
   \label{fig:light curves}
\end{figure}

Fig.~\ref{fig:light curves} shows the $U\!BV\!RI$ light curves of
SN\,2004aw before the seasonal gap. In addition to the results from our
monitoring campaign, an unfiltered measurement provided by T. Boles
and the prediscovery limit of K. Itagaki \citep{IAUC8310} are included
in the figure. The SN maximum is covered in all bands except $U$. The
post-maximum decline in the blue bands is much faster than that in the
redder ones.  The latest points in Fig.~\ref{fig:light curves}
(approximately from day +45 on) mark the onset of the radioactive tail
of the light curve. Unfortunately, several of these magnitudes are
uncertain owing to the very poor S/N (see errors reported in
Table~\ref{SN_mags}).

\begin{figure*}   
   \centering
   \includegraphics[scale=1.0]{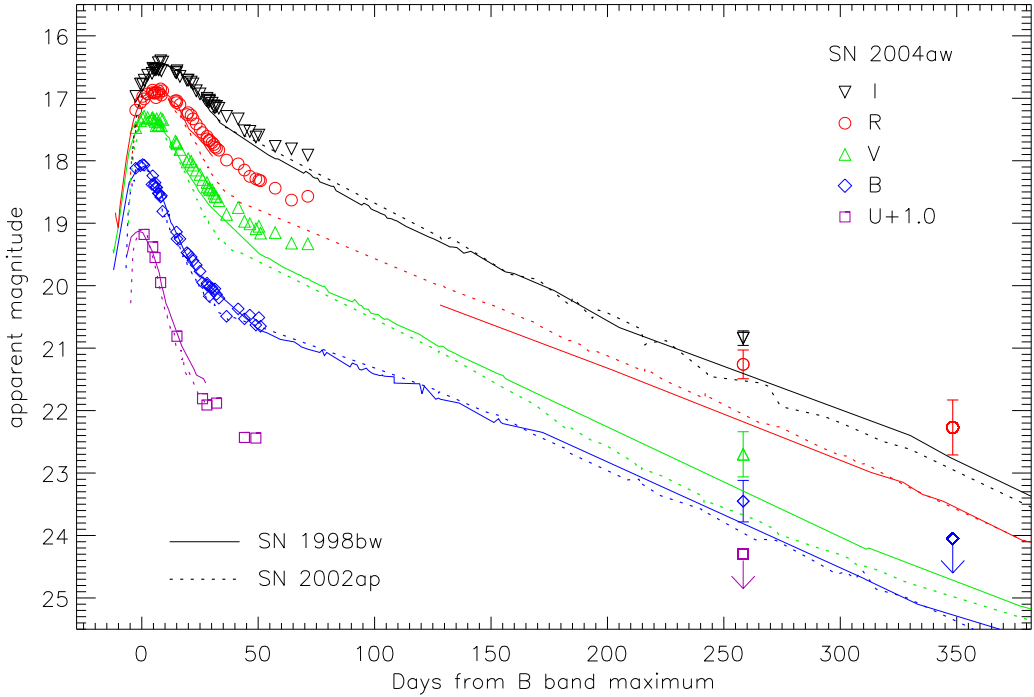}
   \caption{Late-time light curves of SN~2004aw compared to SN~1998bw
\citep{McKenzie99,Patat01} and SN~2002ap
\citep{Foley03,Yoshii03,Tomita06}. The latter two have been shifted in
time and magnitude to match SN~2004aw at maximum. The data suggest a
comparatively slow decline of SN~2004aw.}
   \label{fig:UBVRI_late}
\end{figure*}

Fig.~\ref{fig:UBVRI_late} shows the complete light curves of
SN~2004aw, including the data points obtained at late phases. For
comparison the light curves of SNe~1998bw
\citep{Galama98,McKenzie99,Patat01,Sollerman02} and 2002ap
\citep{Foley03,Yoshii03,Tomita06} are also displayed in the figure,
shifted in time and magnitude to match the peaks of SN~2004aw in all
filters. During the radioactive tail of the light curves, one complete
$U\!BV\!RI$ set of observations was taken 258 days past maximum at
TNG, and in $B$ and $R$ the SN was observed again 348 days past
maximum at the VLT under poor seeing conditions (for details see
Table~\ref{SN_mags}). Since no template is available for these
observations, the non-isolated position of SN~2004aw limits the
precision of our late-time photometry. The SN is not visible in the
TNG $U$-band image and in the VLT $B$-band image, so that only limits
can be derived.

\begin{table*}
\begin{center}
\caption{Main photometric parameters of SN~2004aw.$^a$}
\label{phot-quant}
\begin{footnotesize}
\begin{tabular}{lccccc}
\hline
                & $U$ & $B$ & $V$ & $R$ & $I$\\
\hline
Apparent mag    &  18.18\,$\pm$\,0.15$^b$ &  18.06\,$\pm$\,0.04 &  17.30\,$\pm$\,0.03 &  16.90\,$\pm$\,0.03 &  16.53\,$\pm$\,0.03\\
[-0.3ex]
at max          &                     &                     &                     &                     &                    \\
[0.6ex]
Absolute mag    & $-17.79\,\pm$\,0.56$^b$\hspace{0.2cm} & $-17.63\,\pm$\,0.48\hspace{0.2cm} & $-18.02\,\pm$\,0.39\hspace{0.2cm} & $-18.14\,\pm$\,0.34\hspace{0.2cm} & $-18.18\,\pm$\,0.28\hspace{0.2cm} \\
[-0.3ex]
at max$\,^c$ &                     &                     &                     &                     &                    \\
[0.6ex]
Extinction      &   1.80\,$\pm$\,0.49 &   1.52\,$\pm$\,0.41 &   1.15\,$\pm$\,0.31 &   0.87\,$\pm$\,0.24 &   0.54\,$\pm$\,0.15\\
[1ex] 
Epoch of max    &                     &        0.0         &   +2.7\,$\pm\,$0.6  &   +6.6\,$\pm$\,0.6  &   +8.4\,$\pm\,$0.9 \\
[-0.3ex]
relative to $B\,^d$ &          &                     &                     &                     &                    \\
[0.6ex]
Decline rate $\beta^e$ & 10.28\,$\pm$\,0.60 & 6.96\,$\pm$\,0.16 & 4.64\,$\pm$\,0.12 & 3.20\,$\pm$\,0.12 &   2.16\,$\pm$\,0.12\\[1ex]
Decline rate of &                     &  1.35\,$\pm$\,0.18  &   1.74\,$\pm$\,0.22 &  1.36\,$\pm$\,0.14  &  1.53\,$\pm$\,0.09\\[-0.3ex]
radioactive tail$^f$\\[1ex]
$\Delta m_{15}^g$ &  \hspace{0.05cm} 1.62\,$\pm$\,0.25$\,^b$ &   1.09\,$\pm$\,0.04 &   0.62\,$\pm$\,0.03 &   0.41\,$\pm$\,0.03 &   0.34\,$\pm\,$0.03 \\
\hline
                 & $(U-B)_0$ & $(B-V)_0$ & $(V-R)_0$ & $(R-I)_0$ & $(V-I)_0$\\
\hline
Colour at       &  $-0.16\,\pm$\,0.16$\,^b$ &   0.35\,$\pm$\,0.05 &   0.02\,$\pm$\,0.04 &  $-0.03\,\pm$\,0.04 &  $-0.01\,\pm$\,0.04\\
[-0.3ex]
$B$ maximum$\,^c$ &                   &                     &                     &                     &                    \\
\hline\\[-3.0ex]
\end{tabular}
\end{footnotesize}
\end{center} 
\begin{footnotesize}
$^a$ A distance modulus $\mu$ = 34.17 mag (LEDA, but $H_0=72\,\rmn{km}\,\rmn{s}^{-1}\rmn{Mpc}^{-1}$) and a
colour excess $E(B-V)=0.37$ mag were adopted.\\ 
$^b$ Calculated using the earliest data point in the $U$ band, which
is estimated to be within 2 days from maximum.\\
$^c$ Values corrected for interstellar extinction.\\
$^d$ Based on a polynomial fit and overplotted light curves of other SNe
Ic; the $B$-band maximum is on JD 2\,453\,088.4 $\pm$ 0.5.\\
$^e$ Average decline rate in the time interval 5--30 days past 
$B$-band maximum (in mag/100~d).\\
$^f$ Average decline rate in the time interval 60--300 days past 
$B$-band maximum (in mag/100~d).\\
$^g$ Decline in magnitudes within 15 days from peak.\\
\end{footnotesize}
\end{table*}

In Table~\ref{phot-quant} the most important photometric properties of
SN~2004aw are reported. One of them is the clear delay of maximum
light in the red with respect to the bluer bands, which is highlighted
by a dotted line in Fig.~\ref{fig:light curves}. Between the maxima in
the $B$ and $I$ bands, for instance, there is an offset of 8.4
days. Fig.~\ref{fig:UBVRI_late} reveals that the early $B$-band light
curve of SN~2004aw (until $\sim$\,40 days after the $B$-band maximum)
overlaps with that of SN~1998bw. The $U$ band seems to fade slightly
faster (very similar to SN~2002ap), while in the redder bands the
decline of SN~2004aw is slower than that of SN~1998bw.  Even with
measurement uncertainties of up to 0.4 mag (see Table~\ref{SN_mags}), the
late-time light curves of SN~2004aw shown in Fig.~\ref{fig:UBVRI_late}
seem to deviate from those of the other two SNe. Measured relative to
the peak, SN~2004aw appears between 0.4 mag ($B$~band) and 0.9 mag
($R$~band) more luminous than SN~1998bw at day +258, and a comparison
with SN~2002ap yields similar results. This can be interpreted either
as a measurement error due to a possible underlying compact source of
light in the host galaxy, or as a real effect, making SN~2004aw a very
slowly declining SN. The fact that even in the nebular spectra taken
at similar epochs (Table~\ref{spectra}), almost no trace of an
underlying stellar continuum or narrow H$\alpha$ emission can be
detected (see Section~\ref{Sequence of optical spectra}), indicates
that the galaxy background at the site of SN~2004aw is smooth enough
to be properly subtracted. Hence, we tend to believe in our
measurements and think that SN~2004aw truly declines more slowly than
SNe~1998bw and 2002ap.

\subsection{Near-IR light curves}
\label{JHK photometry}

To date our knowledge of the behaviour of SNe~Ic in the near-IR
is very limited. Only a few objects have ever been observed in the IR,
and in most cases the coverage is rather poor. IR photometry at more
than $\sim$\,5 epochs is available only for SN~2002ap
\citep{Nishihara_proc,Yoshii03}. Given the paucity of IR observations,
even our modest IR dataset of SN~2004aw may shed some light on the
behaviour of SNe~Ic in this wavelength range.

\begin{figure}   
   \centering
   \includegraphics{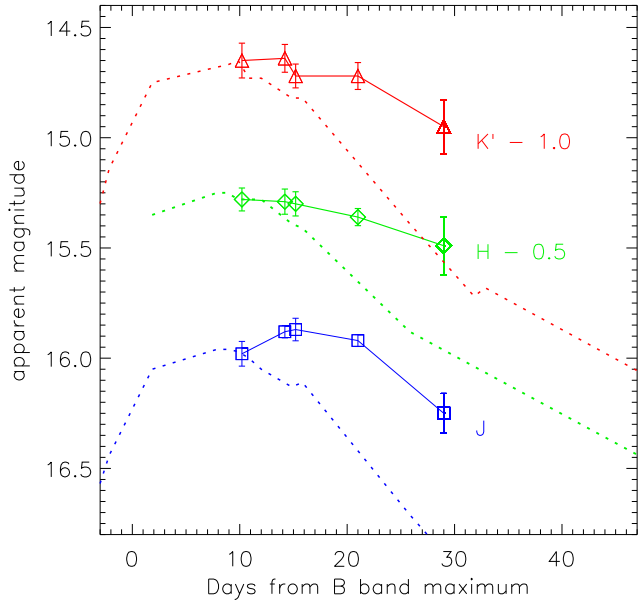}
   \caption{$JHK'$ light curves of SN~2004aw. For comparison the $JHK$
light curves of SN~2002ap are also displayed \citep[dotted
lines,][]{Yoshii03}. A global shift, besides the one mentioned in the
figure for each band, has been applied to the data of SN~2002ap to
roughly match SN~2004aw at +10 days.}
   \label{fig:JHK}
\end{figure}

As mentioned in Section~\ref{Photometry}, no colour corrections were
applied to calibrate the $JHK'$ magnitudes of SN~2004aw presented in
Table~\ref{IR_mags}. This is a considerable source of uncertainty;
moreover, the sparse sampling makes it difficult to check the
self-consistency of the light curves, as can be seen in
Fig.~\ref{fig:JHK}. The $H$ and $K'$ light curves show a relatively
flat behaviour with only a slight decline ($\leq 0.3$ mag) in the
interval from +10 to +30 days. The $J$ band seems to peak at about day
+17, but we cannot exclude that because of lacking colour corrections
the calibration error of some points is actually larger than that reported
in Table~\ref{IR_mags}.  During the time interval covered by our
near-IR photometry, all optical bands show a significant decline, from
about 0.6 mag in the $I$ band to 1.7 mag in the $U$ band. The
different behaviour of the optical and the near-IR bands in this phase
is responsible for the strongly increasing contribution of the near-IR
to the total bolometric flux, which is discussed in
Section~\ref{Absolute magnitudes and bolometric light curve}.

To date SN~2002ap is the only SN~Ic with good temporal coverage
of the near-IR bands at early phases \citep{Nishihara_proc,Yoshii03},
making a systematic comparison of SNe~Ic in this regime
difficult. At least between day +10 and day +30, SN~2002ap and
SN~2004aw show substantial differences in their evolution, as can be
seen in Fig.~\ref{fig:JHK}. While SN~2004aw exhibits an almost
plateau-like behaviour, the $JHK'$ light curves of SN~2002ap fade with
rates similar to the optical bands, ranging from 0.75 to 1.0 mag
within these 20 days. Given the fairly similar optical light curves of
these two SNe, the differences in the near-IR are surprising.

\subsection{Colour evolution}
\label{Colour evolution}

\begin{figure*}   
   \centering
   \includegraphics{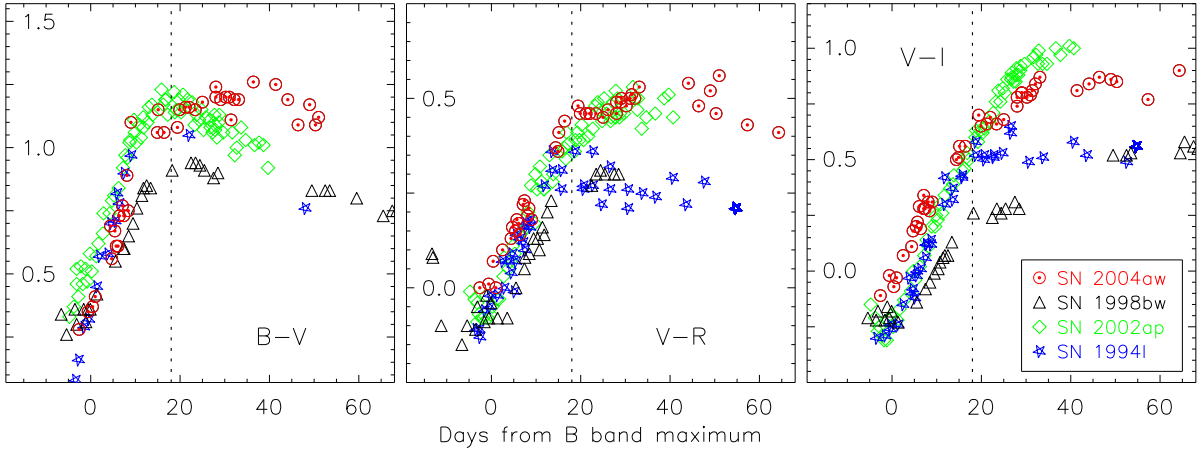}
   \caption{Colour evolution of SNe 2004aw, 1998bw, 2002ap, and
1994I. The curves have been dereddened according to the values
reported in Section~\ref{Absolute magnitudes and bolometric light
curve} (in particular $E(B-V)=0.30$ mag for SN~1994I) and shifted in
time to match the $B$-band maximum of SN~2004aw. The vertical dotted
lines at +18 days mark the epoch suggested to be used for dating SNe
Ic (see Section~\ref{Colour evolution}).}
   \label{fig:04aw_98bw_02ap_94I_colors}
\end{figure*}

Fig.~\ref{fig:04aw_98bw_02ap_94I_colors} presents the evolution of the
intrinsic $B-V$, $V-R$, and $V-I$ colours for SNe 2004aw, 1998bw,
2002ap, and 1994I. The extinction values reported in
Sections~\ref{Reddening and distance} and \ref{Absolute magnitudes and
bolometric light curve} were used to deredden the colour
curves. Starting before maximum light, the colours of all objects
become monotonically redder for more than 2 weeks. Later on, the
differences among the different objects become more distinct.

The $B-V$ colours of SN~1994I and SN~2002ap are reddest at about +15
to +20 days, and then become bluer again. This also holds for
SN~1998bw, although here the evolution after the red peak is much
slower. In contrast, in SN~2004aw this epoch marks the onset of a
plateau of fairly constant $B-V$ colour lasting at least until +50
days. During this plateau phase SN~2004aw is typically 0.3 to 0.4 mag
redder than SN~1998bw.  The $V-R$ colour evolution of SN~2004aw and
SN~2002ap is very similar, whereas SN~1994I becomes increasingly bluer
after day +18. SN~1998bw shows the bluest $V-R$ colour until +15 days,
but the subsequent behaviour cannot be evaluated due to lack of
information in the $R$ band after +30 days. At +30 days, SN~2004aw and
SN~2002ap are redder by $\sim$\,0.2 mag than SN~1994I and SN~1998bw.
Finally, SN~2004aw shows the reddest $V-I$ colour until day +18. After
this epoch, SN~1998bw and SN~2004aw still become redder, but with a
reduced slope. SN~2002ap takes another 10 days to reach this
inflection point, while the $V-I$ colour of SN~1994I remains constant
between +20 and +50 days. At +30 days there is a clear colour
separation between the objects with a $V-I$ of 0.3, 0.5, 0.8, and 0.9
mag for SN~1998bw, SN~1994I, SN~2004aw, and SN~2002ap, respectively.

In all the colours studied here, SN~1998bw appears to be bluer than
SN~2002ap and SN~2004aw during the entire post-maximum phase. After day
+20, the same holds for SN~1994I, but in this case this may be due to
contamination from blue stars in the host galaxy (see also Section~
\ref{Spectroscopic comparison with other SNe Ic}). Remarkably,
the SNe of our sample all show a peak or a change in the slope of
their colour curves at about 15 to 20 days after $B$-band maximum. If
the generality of such behaviour can be confirmed using a
significantly larger sample of well-monitored SNe~Ic, this may provide
an interesting, extinction-independent new tool to determine the phase
of SNe~Ic and BL-SNe when maximum light was missed by not more
than a few days.

\subsection{Absolute magnitudes, bolometric light curve}
\label{Absolute magnitudes and bolometric light curve}
 
In order to compare the absolute peak magnitudes, the values reported
in the literature should first be homogenised with respect to the
treatment of distance and extinction. Whenever available, we preferred
a Cepheid distance modulus. When a kinematical distance modulus had to
be used, it was computed using the host-galaxy recession velocity
corrected for the Local-Group infall onto the Virgo cluster and
$H_0=72\,\rmn{km}\,\rmn{s}^{-1}\rmn{Mpc}^{-1}$. In most cases,
extinction estimates from the original papers were adopted, but for
SN~1994I a more recent result provided by \citet{Sauer06} was used.

Comparing the absolute magnitudes, SN~2004aw seems to be a fairly
bright, but not outstanding object. Type Ic SNe exhibit peak
magnitudes in the $V$ band from about $-17$ to beyond $-19$ (see,
e.g., \citealt{Richardson06}). With M$_V=-19.13$ mag
(\citealt{Galama98}, $\mu=32.76$ mag, $E(B-V)=0.06$ mag), the
BL-SN~1998bw is one of the brightest core-collapse SNe ever
observed. The broad-lined SN~2002ap, on the other hand, is
comparatively faint at maximum with M$_V =-17.35$ mag
\citep[][$\mu=29.46$ mag, $E(B-V)=0.09$
mag]{Sharina96,Foley03,Yoshii03}, and with M$_V$ = $-17.14$ mag, the
BL-SN~1997ef is not very luminous either \citep[][$\mu=33.54$ mag,
$E(B-V)=0.04$ mag]{IAUC6778,IAUC6786,IAUC6798,Mazzali02}. The peak
absolute magnitudes of the so-called ``normal" SNe~Ic seem to cover a
similarly wide range. SN~1994I peaks at M$_V = -17.62$ mag
\citep[][$\mu=29.60$ mag, $E(B-V)=0.30$ mag]{Richmond96,Sauer06},
SN~1999ex reaches $-17.58$ mag \citep[][$\mu=33.28$ mag, $E(B-V)=0.30$
mag]{Stritzinger02}, and SN~1992ar may have M$_V < -20$ mag
\citep{Clocchiatti99}. SN~2004aw (M$_V$ = $-18.02\ \pm$ 0.39 mag) is a
bit brighter than SN~1994I (and probably also the average of all
SNe~Ic), but not exceptionally luminous.

\begin{figure}   
   \centering
   \includegraphics[scale=1.0]{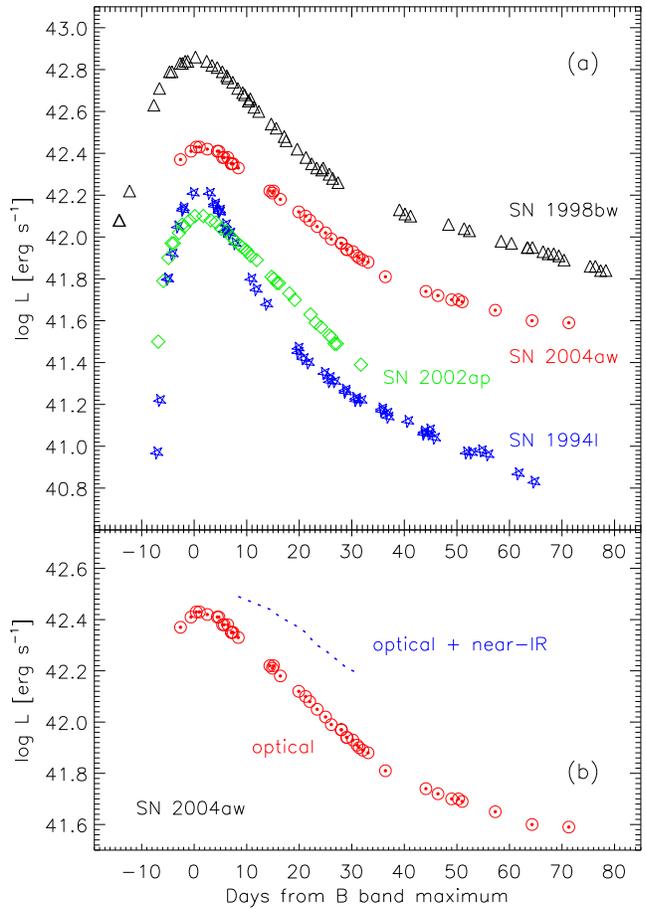}
   \caption{(a) Quasi-bolometric light curves of SN~2004aw, SN~1998bw,
SN~2002ap, and SN~1994I, computed including the $U$-through-$I$-band
fluxes. The distance and extinction values reported in
Section~\ref{Absolute magnitudes and bolometric light curve} have been
adopted.
\newline
   (b) Comparison of the quasi-bolometric light curves of SN~2004aw
with (dotted line) and without the contribution of the $JHK'$
bands. For phases earlier than +10 days and later than +30 days no
near-IR photometry is available.}
   \label{fig:bolom}
\end{figure}

We have constructed a ``quasi-bolometric light curve'' \citep[see,
e.g.,][]{Nomoto90} of SN~2004aw in the following way: first the $U$
through $I$ magnitudes were converted into monochromatic fluxes and
the spectral energy distribution (SED) was interpolated linearly. Then
the SED was integrated over frequency, assuming zero flux at the
integration limits, which are given by the blue edge of the $U$ band
and the red edge of the $I$ band.  In Fig.~\ref{fig:bolom}(a) we
compare the bolometric curve of SN~2004aw with those of SN~1998bw,
SN~2002ap and SN~1994I. We note a wide spread in peak luminosities,
with SN~1998bw being the brightest object and SN~2004aw
intermediate. The differences in the decline rates observed in
different bands are clearly reproduced in the bolometric light curves.

Taking into account the contribution of the $JHK'$ bands (see
Section~\ref{JHK photometry} and Table~\ref{IR_mags}) makes a
significant difference in the bolometric light curves, as can be seen
in Fig.~\ref{fig:bolom}(b). Between +10 days and +30 days the near-IR
contribution increases from $\sim$\,31\% to $\sim$\,45\%,
demonstrating that a considerable portion of the bolometric flux is
released in these bands. Unfortunately, no near-IR photometry is
available for epochs earlier than +10 days or later than +30 days, so
that the corrections to the $U$-through-$I$ light curve cannot be
obtained for all phases. \citet{Yoshii03} estimated that in SNe~Ic,
wavelength regions other than the optical and near-IR contribute
much less than 10\% to the total bolometric flux. Therefore, ignoring
the contributions from these frequencies does not lead to a
significant underestimate of the true bolometric luminosity.

\begin{figure*}  
   \centering
   \includegraphics[scale=1.0]{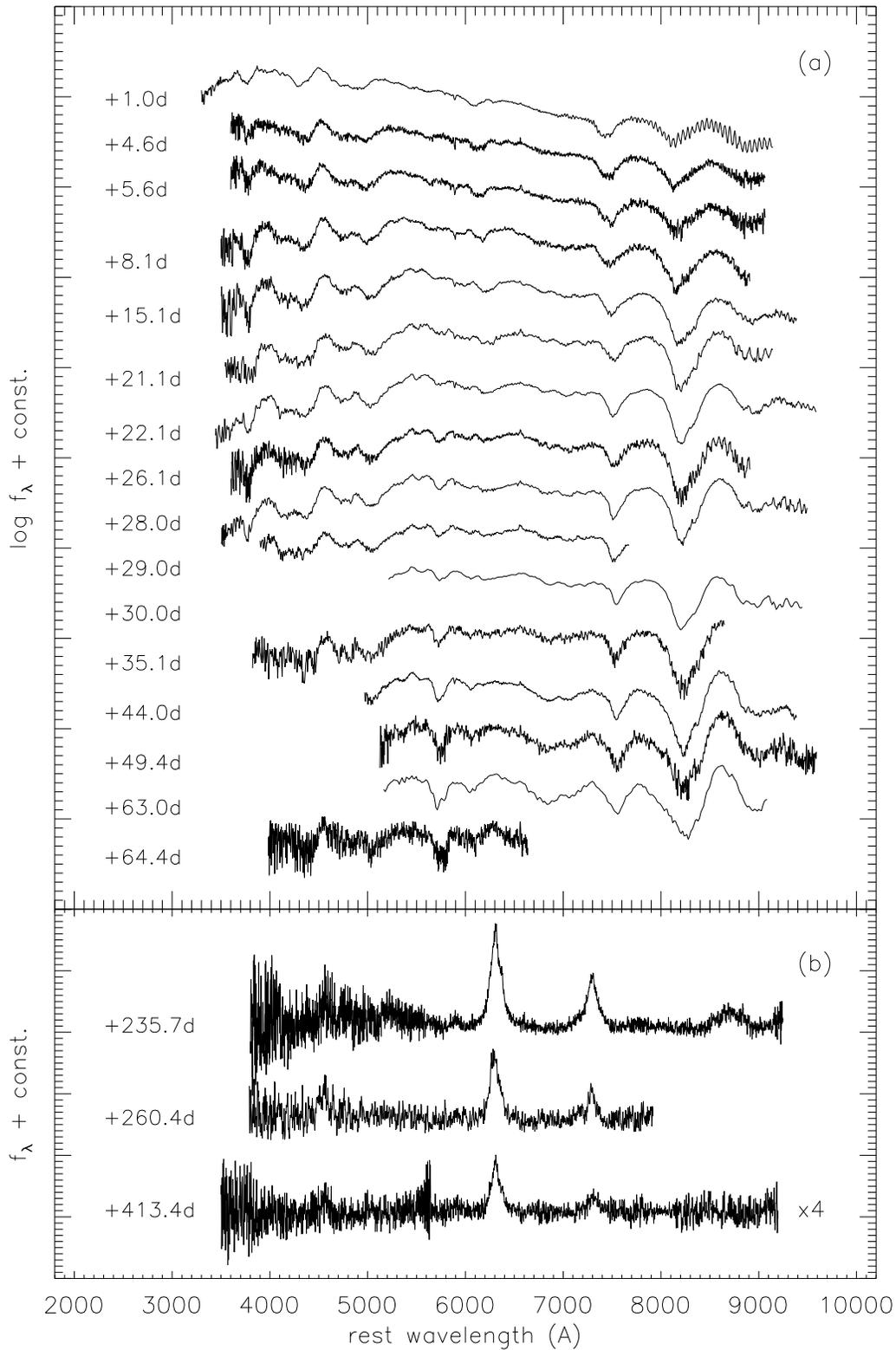}
   \caption{Spectroscopic evolution of SN~2004aw. Epochs are given
with respect to $B$-band maximum (JD = 2\,453\,088.4). The spectra were
checked against the photometry, transferred to the SN rest frame
assuming a recession velocity of 4900 km\,s$^{-1}$ (inferred from the
position of a faint host galaxy H$\alpha$ line visible in most
spectra), and dereddened assuming $E(B-V)$ = 0.37 mag.
\newline
   (a) Series of photospheric spectra from maximum light to about 2
months later, shown with a logarithmic scale. The day +5.6 spectrum is a
combination of the two spectra taken at Siding Spring and Asiago
within half a day. 
\newline
   (b) Nebular spectra, plotted linearly. For better clarity the
day +413 spectrum was scaled up by a factor of 4. The increased noise
in this spectrum around 5500~\AA\ marks the overlap region of the blue 
and red channels of Keck/LRIS, with low signal in both of them. }
   \label{fig:spectra}
\end{figure*}

\section{Spectral evolution}
\label{Spectral evolution}

\subsection{Sequence of optical spectra}
\label{Sequence of optical spectra}

The spectroscopic evolution of SN~2004aw from maximum light to +65
days is densely covered. During this period a continuum is present in
all the spectra, so that the photospheric phase lasts for at least two
months from maximum. However, after approximately one month the
emission component of the Ca\,{\sc ii} near-IR triplet starts to grow,
and in the last two spectra of this sequence first hints of forbidden
O\,{\sc i} and Ca\,{\sc ii} emission lines become visible, marking the
onset of the transition to the nebular phase. In addition, three
entirely nebular spectra at epochs between 236 and more than 400 days
past maximum have been obtained. The complete spectroscopic evolution
is presented in Fig.~\ref{fig:spectra}.

\begin{figure}   
   \centering
   \includegraphics{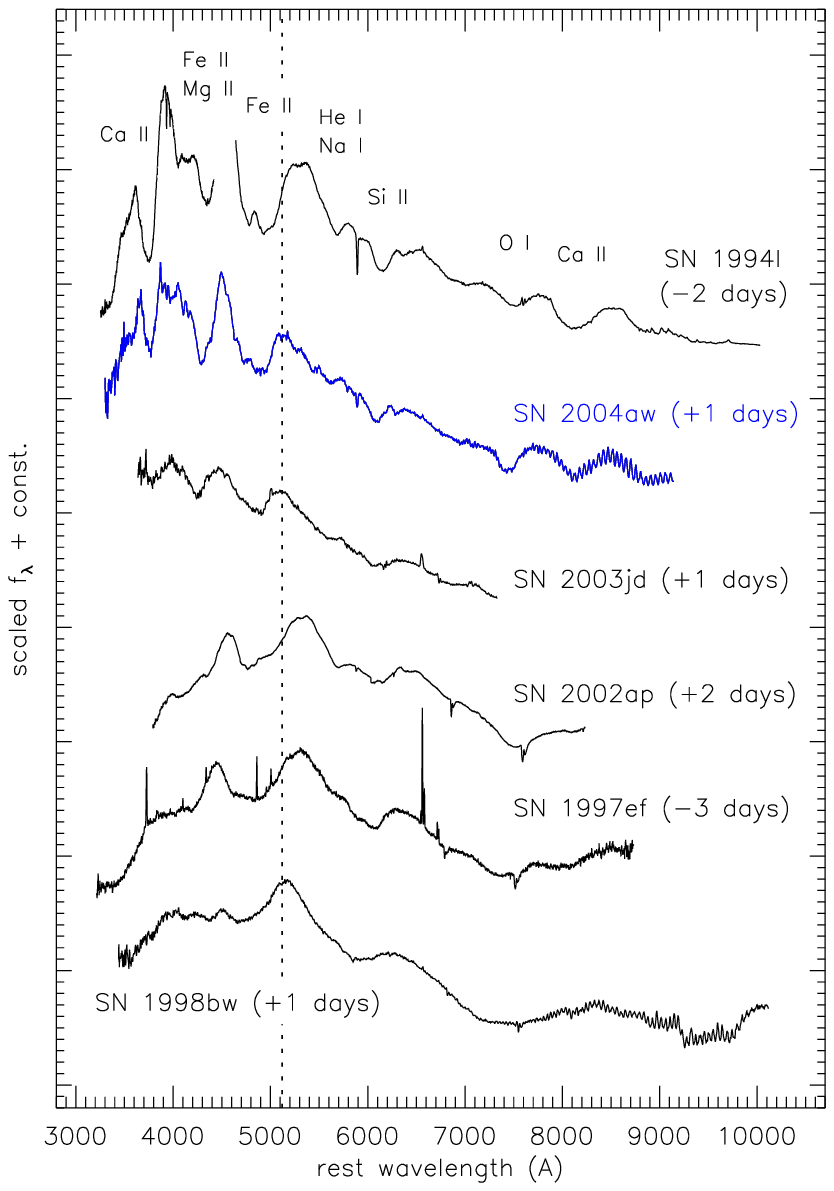}
   \caption{Comparison of spectra of SNe~Ic close to maximum
light. The spectra have been dereddened assuming $B-V$ colour excesses
of 0.30 mag for SN~1994I \citep{Sauer06}, 0.37 mag for SN~2004aw, 0.14
mag for SN~2003jd (Valenti et al., in preparation), 0.04 mag for
SN~1997ef \citep[only Galactic extinction,][]{Schlegel98}, 0.09 mag
for SN~2002ap \citep{Yoshii03}, and 0.06 mag for SN~1998bw
\citep{Galama98}. The order of plotting suggests increasing ejecta
velocities from top to bottom.}
   \label{fig:comp_plus01}
\end{figure}

\begin{figure}   
   \centering
   \includegraphics{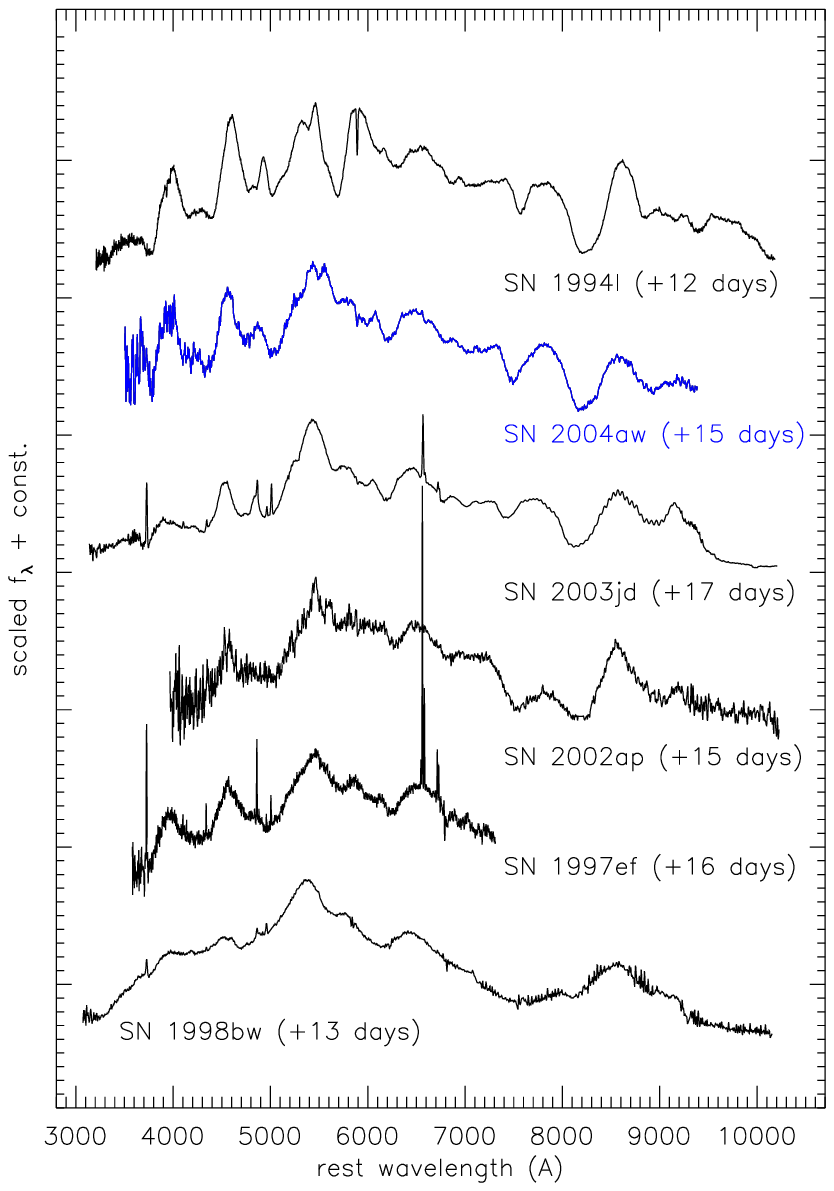}
   \caption{The same as Fig.~\ref{fig:comp_plus01}, but 2 weeks after
maximum light.}
   \label{fig:comp_plus15}
\end{figure}

\begin{figure}   
   \centering
   \includegraphics{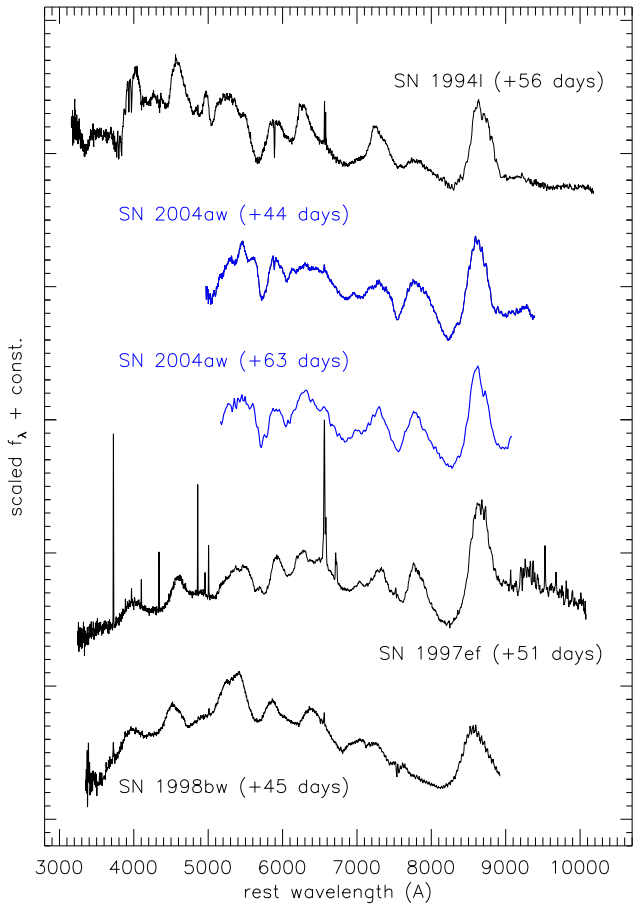}
   \caption{The same as Fig.~\ref{fig:comp_plus01}, but 6 to 9 weeks
after maximum light.}
   \label{fig:comp_plus44}
\end{figure}

\begin{figure}   
   \centering
   \includegraphics{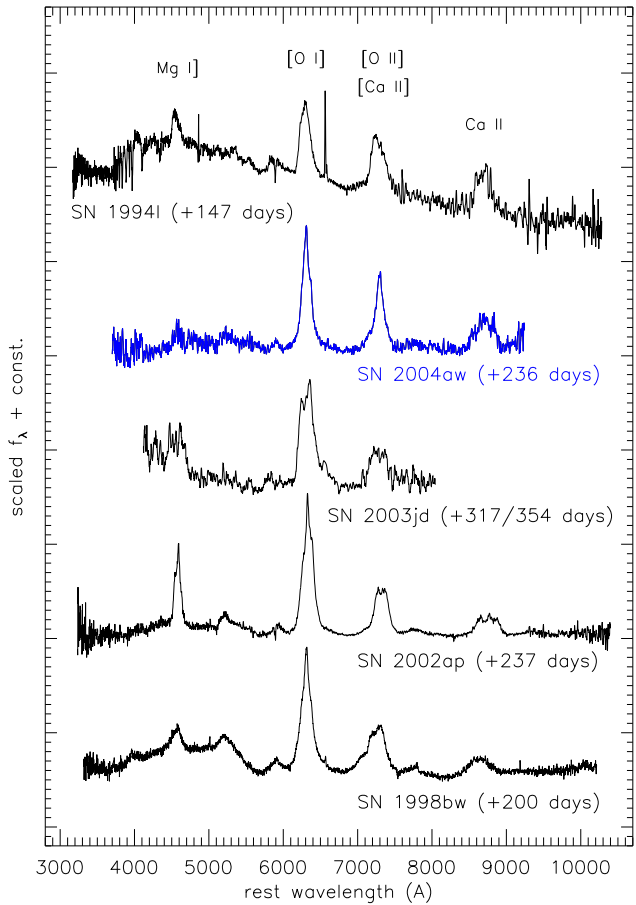}
   \caption{The same as Fig.~\ref{fig:comp_plus01}, but 5 to 12 months
after explosion. The spectrum of SN~2004aw was smoothed with a box
size of 10~\AA. The spectrum of SN~2003jd is a combination of two
spectra taken 317 and 354 days after maximum, respectively. SN~1994I
is heavily contaminated by light from the underlying galaxy.}
   \label{fig:comp_plus236}
\end{figure}

The first spectrum of this sequence, obtained one day after $B$-band
maximum, is characterised by a blue continuum and P-Cygni features of
Ca\,{\sc ii} (H\&K and the near-IR triplet), O\,{\sc i} $\lambda7774$,
Fe\,{\sc ii} blends, and Mg\,{\sc ii} $\lambda4481$. Also, Si\,{\sc ii}
$\lambda6355$ and possibly C\,{\sc ii} $\lambda6580$ and Na\,{\sc
i}\,D can be seen. He\,{\sc i} $\lambda5876$ might be blended with
Na\,{\sc i}\,D, although other optical He\,{\sc i} lines cannot be
detected. During the first week after maximum (covered by the first four
spectra) the evolution is significant. Within a few days the flux in
the blue part decreases dramatically, the Ca\,{\sc ii} near-IR triplet
becomes stronger, and the absorption troughs centred at 4200~\AA\ and
4800~\AA\ develop the characteristic "W\,"-\,shaped profile that is
observed in many Type I SNe around and after maximum. 

Later on the spectra evolve more slowly, and during the following
month the main changes consist of a further suppression of the flux in
the blue, an increasing strength of the emission component of the
Ca\,{\sc ii} near-IR triplet, and an increasingly distinct Na\,{\sc
i}\,D absorption. The Si II line, on the other hand, disappears
(\citealt{Clocchiatti97} mentioned the possibility that it might be
filled by the emerging [O\,{\sc i}] $\lambda\lambda6300,6364$ emission
lines). Toward the end of our early-time follow-up, approximately 65
days after maximum, SN~2004aw still shows a well-developed continuum,
with only hints of the characteristic nebular emission lines
superimposed.

About 170 days later (at a phase of +236 days) the continuum has
entirely disappeared, and the spectrum is dominated by forbidden
emission lines of [O\,{\sc i}] $\lambda\lambda6300,6364$ and [Ca\,{\sc
ii}] $\lambda\lambda7291,7323$, possibly blended with [O\,{\sc ii}]
$\lambda\lambda7320,7330$. Emissions of the Ca\,{\sc ii} near-IR
triplet, Mg\,{\sc i}] $\lambda4571$ and blended [Fe\,{\sc ii}] lines
near 5000~\AA\ are also visible. While the weak feature near 6000~\AA\
could be a residual of Na\,{\sc i}\,D, the faint emission line near
4000~\AA\ can tentatively be attributed to [S\,{\sc ii}]
$\lambda4069$. The peaks of both [O\,{\sc i}]
$\lambda\lambda6300,6364$ and [Ca\,{\sc ii}] $\lambda\lambda7291,7323$
/ [O\,{\sc ii}] $\lambda\lambda7320,7330$ are sharp and show a narrow
core. Between 236 and 413 days past maximum only minor changes
occurred, the most obvious being the disappearance of the permitted
emission of the Ca\,{\sc ii} near-IR triplet.  The strength of the
[Ca\,{\sc ii}] relative to [O\,{\sc i}] also declines.

\subsection{Comparison to optical spectra of other SNe Ic}
\label{Spectroscopic comparison with other SNe Ic}

The spectral evolution of SN~2004aw is consistent with that of other
Type Ic SNe. However, the spectroscopic variety among the members of
this class is large, especially when also the broad-lined supernovae
are taken into account. It is the aim of this section to examine the
differences among SNe Ic with particular attention to SN~2004aw.

Figs. \ref{fig:comp_plus01} to \ref{fig:comp_plus44} show a comparison
of the spectra of SN~2004aw with those of SN~1994I
\citep{Filippenko95} and the BL-SNe 1998bw \citep{Patat01}, 2002ap
\citep{Kawabata02,Foley03}, 1997ef \citep{Mazzali00,Matheson01}, and
2003jd (Valenti et al., in preparation) at maximum light, 2 weeks
later, and 6 to 9 weeks past maximum, respectively. At all three epochs
SN~1998bw is an outstanding object, exhibiting comparatively broad and
shallow features and strong line blending due to a large amount of
ejected material at extremely high velocities. SNe~2002ap and 1997ef
(and to a lesser degree SN~2003jd) are similar to SN~1998bw at early
epochs (Fig.~\ref{fig:comp_plus01}), but then evolve towards more
typical (i.e., SN~1994I-like) Type Ic spectra (see
Fig.~\ref{fig:comp_plus44}). Compared with SNe~1994I and 2004aw around
maximum light, SNe~1998bw, 1997ef and 2002ap show a clear flux deficit
shortward of $\sim$\,5000 \AA\ due to strong line blanketing. Worth
mentioning is also the difference in the exact position of the broad
emission feature lying between 5000 \AA\ and 5500 \AA\ in all
maximum-light spectra. In SNe~2004aw, 2003jd and 1998bw its peak is
located at $\sim$\,5120 \AA\ (indicated by a vertical line in
Fig.~\ref{fig:comp_plus01}), whereas in SNe~1994I, 1997ef and 2002ap
it is shifted by about 200~\AA\ to the red.

\begin{figure*}   
   \centering
   \includegraphics{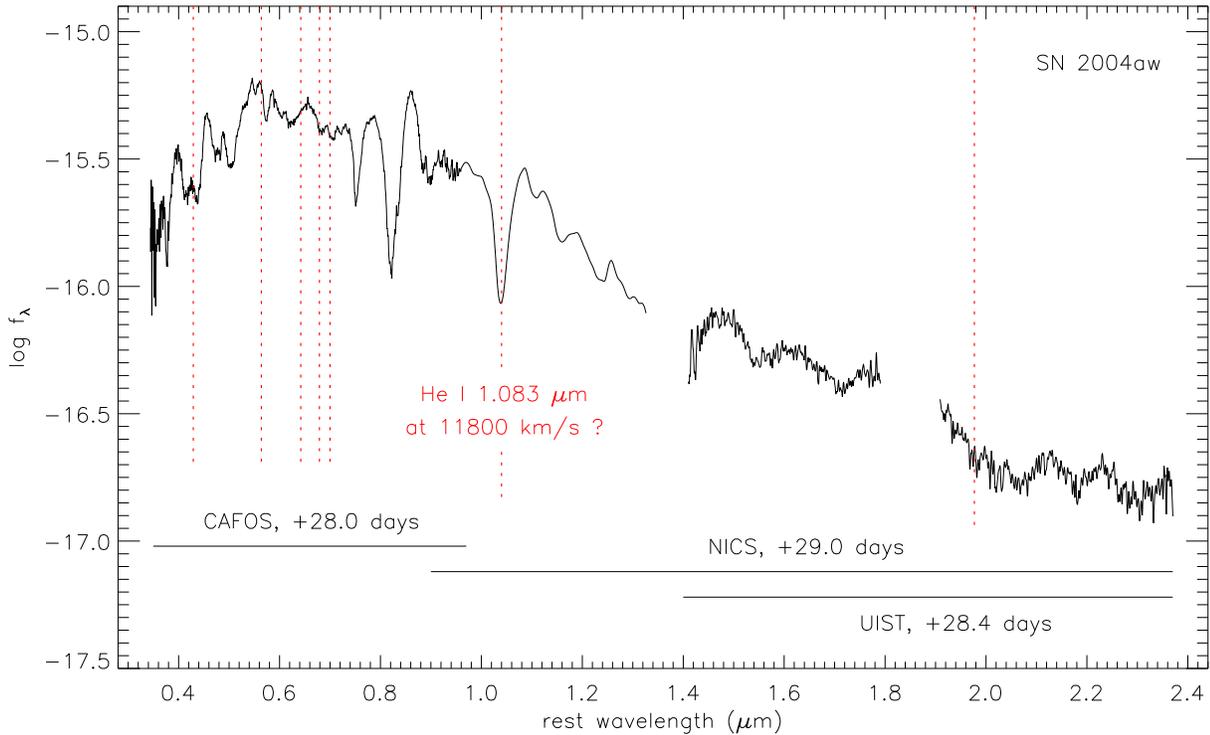}
   \caption{Combined optical and near-IR spectrum of SN~2004aw taken
four weeks after $B$-band maximum. In the bottom of the figure the ranges
of the three individual spectra are indicated. Vertical dotted lines
mark the expected position of strong He\,{\sc i} features for an
expansion velocity of 11\,800 km\,s$^{-1}$.}
   \label{fig:IR_opt_spec}
\end{figure*}

By +12 days (see Fig.~\ref{fig:comp_plus15}) Na\,{\sc i}\,D (possibly
blended with He\,{\sc i}, see \citealt{Filippenko95,Clocchiatti96}) is
one of the dominant features in the spectrum of SN~1994I, while it is
still rather weak in all the other objects in the figure. At similar
epoch it is also weak in SNe~1983V \citep{Clocchiatti97} and 1987M
\citep{Filippenko90} which are not shown here. Moreover, the emerging
nebular emission features of [O\,{\sc i}] $\lambda\lambda6300,6364$
and [Ca\,{\sc ii}] $\lambda\lambda7291,7323$ are clearly seen in
SN~1994I at +56 days, while at best only a hint of them can be
detected in the spectra of the other SNe at comparable phases (see
Fig.~\ref{fig:comp_plus44}). This suggests that SN~1994I underwent
relatively fast spectral evolution.

The spectrum of SN~2004aw shows similarities to both SN~1994I and the
group of moderate BL-SNe represented by SNe~1997ef, 2002ap, and
2003jd. At maximum (Fig.~\ref{fig:comp_plus01}) the width of the lines
of SN~2004aw is well matched by SN~1994I and is significantly smaller
than in all BL-SNe except SN~2003jd. However, SN~2004aw does not
follow the fast evolution of SN~1994I, and hence at later times
(Figs.~\ref{fig:comp_plus15} and \ref{fig:comp_plus44}) it bears a
closer resemblance to the spectra of the BL-SNe~2003jd and
1997ef. Also, the Na\,{\sc i}\,D feature is much weaker than in
SN~1994I at all epochs.

All nebular spectra shown in Fig.~\ref{fig:comp_plus236} have in
common that [O\,{\sc i}] $\lambda\lambda6300,6364$ is the strongest
feature. However, the shape of this feature reveals important
differences: in SNe~2004aw, 1998bw, and 2002ap the line has a sharp,
narrow core, whereas in SN~1994I its top is somewhat rounded. In
SN~2003jd the feature even exhibits a double-peaked structure. The
impact that these differences have on our picture of the explosions is
discussed in Section \ref{Discussion}. In SN~2004aw, the [Ca\,{\sc ii}]
$\lambda\lambda7291,7323$ / [O\,{\sc ii}] $\lambda\lambda7320,7330$
emission shows a sharp peak, in contrast to the other SNe where the
top of this feature is flat or rounded. Mg\,{\sc i}] $\lambda4571$ is
particularly strong in SN~2002ap, and rather weak in SN~2004aw. The
[Fe\,{\sc ii}] lines near 5000~\AA\ are more prominent in SN~1998bw
than in other SNe of this sample.

\subsection{Near-IR spectroscopy: Identification of He\,{\sc i} lines?}
\label{Near-IR spectroscopy}

Few SNe~Ic have been followed spectroscopically in the IR. Only for
two of them are the data public: SN~1999ex \citep[actually an object
intermediate between Type Ib and Ic;][]{Hamuy02} and the highly
energetic BL-SN~1998bw \citep{Patat01}. However, near-IR spectroscopy
may turn out to be a powerful tool for investigating important
properties of the progenitor stars of stripped-envelope core-collapse
SNe. Although it is commonly accepted that SNe~Ic originate from
massive stars that have lost most of their H and He envelope, several
questions (e.g., concerning the exact mechanism of mass loss) are yet
unresolved.  Some of these issues may be effectively addressed by a
possible detection of He in the spectra of SNe normally classified as
Type Ic. Since the He\,{\sc i} features in the near-IR are much
stronger than those in the optical regime, IR spectroscopy may play a 
key role.

\begin{figure*}   
   \centering
   \includegraphics{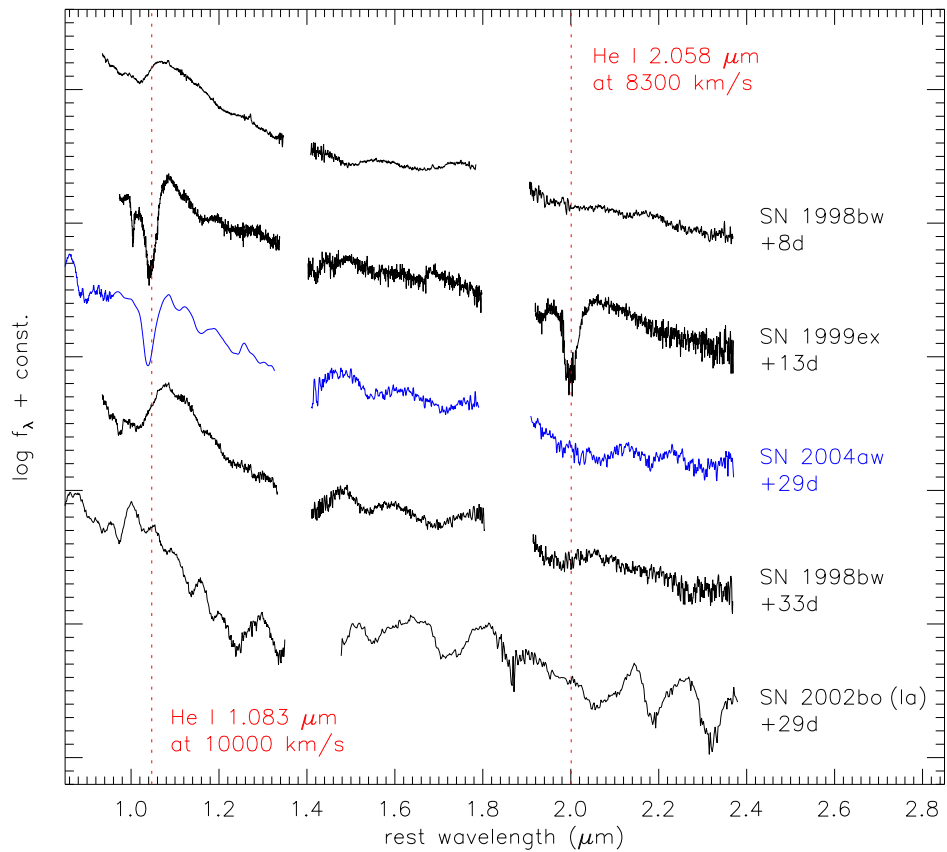}
   \caption{Near-IR spectra of SNe 2004aw (Type Ic), 1999ex (Type
Ib/c), 1998bw (BL-SN), and 2002bo (Type Ia) at epochs from 1 to 5 weeks
after $B$-band maximum. The position of the minima of two pronounced
He\,{\sc i} features in SN~1999ex is marked by vertical dotted
lines. For references see text.}
   \label{fig:IR_spec_comp_Ia}
\end{figure*}

Fig.~\ref{fig:IR_opt_spec} shows a combination of an optical and two 
IR spectra of SN~2004aw $29$ days past maximum, covering the spectral range
0.35\,--\,2.40~$\mu$m. Among the most prominent features is an absorption
at 1.04~$\mu$m. The origin of this absorption, which is also seen in
other SNe~Ic, has been the subject of some debate. For SN~1994I this
feature was originally attributed to He\,{\sc i} \citep{Filippenko95,
Clocchiatti96}. However, \citet{Baron99} disfavoured an identification
with He\,{\sc i} because it would require other optical lines from
this ion to show up in the spectrum as well, which is not seen.
\citet{Millard99} discussed alternative origins for this feature such as
Si\,{\sc i} and C\,{\sc i}, based on line identifications from the
highly parameterised spectral synthesis code {\sc synow}.  

\citet{Sauer06} found that the absorption might be caused by a mix of
some He\,{\sc i} and C\,{\sc i}.  The consistent simulation of
He\,{\sc i} requires nonthermal excitation and ionisation processes
from fast electrons caused by Compton scattering of $\gamma$-ray
photons from $^{56}$Ni and $^{56}$Co decay.  Without those processes,
the excitation of He\,{\sc i} at the temperatures present in the
ejecta is too low to cause any significant absorption \citep{Lucy91}.
\citet{Sauer06} use a simple estimate of these excitations by
increasing the optical depth in He\,{\sc i} by hand. Although this
method does not provide a self-consistent description of the formation
of the helium lines, their estimate shows that it is unlikely that
He\,{\sc i} alone can form such a broad absorption without showing a
trace of other lines of this ion.

In contrast to SN~1994I, where no observation of the IR wavelength
bands is available, SN~2004aw provides the opportunity to place a
tighter constraint on the contribution of the He\,{\sc i} $1.083$~$\mu$m
line to the $1.04$~$\mu$m absorption, because the presence of
significant amounts of He\,{\sc i} should be detectable in the
infrared through the $2.058$~$\mu$m line. The position of all strong
He\,{\sc i} lines at a velocity of $11800\,$km\,s$^{-1}$ is indicated
in Fig.~\ref{fig:IR_opt_spec} by vertical dotted lines.  There is no
clear detection of a line at $\sim\,1.98$~$\mu$m, which provides strong
evidence that the contribution of He\,{\sc i} to the $1.04$~$\mu$m
absorption must indeed be small. In contrast to other helium lines, the
wavelength region of this strong IR line is not crowded with lines
from other ions.  \citet{Mazzali98} studied the formation of the
proposed He\,{\sc i} line in the spectra of the Type~Ia SN~1994D using
a full non-LTE description of this ion including the nonthermal
excitations. They found that He\,{\sc i} $2.058$~$\mu$m clearly showed
up as the second strongest line in their model spectra.  Therefore, a
detection should be possible if He\,{\sc i} alone is responsible for
the $1.04$~$\mu$m feature.  Based on IR observations of SNe~2000ew,
2001B, and 2002ap, \citet{Gerardy_proc} reached a similar conclusion.

\citet{Patat01} dealt with the formation of the near-IR He\,{\sc i}
features in great detail, including a discussion of factors (intrinsic
line strengths, level populations, non-thermal excitation, resonance
scattering enhancement of the He\,{\sc i} $1.083$~$\mu$m line) which
may influence the ratio of the EWs of He\,{\sc i} $2.058$~$\mu$m to
He\,{\sc i} $1.083$~$\mu$m (hereafter $\mathcal{R}_{\rmn{He}}$). They
conclude that a $\mathcal{R}_{\rmn{He}}$ of less than 1 is not
unexpected, consistent with a putative detection of He\,{\sc i} in the
near-IR spectra of SN~1998bw at an observed line-width ratio of
$\sim1/3$. 

However, values of $\mathcal{R}_{\rmn{He}}$ close to 1 seem to be
feasible, as confirmed by SN~1999ex, an intermediate case between
SN~Ib and SN~Ic with weak He\,{\sc i} lines visible in the optical
part of the spectrum. For this SN three near-IR spectra were obtained
by \citet{Hamuy02}. The latest one, at an epoch of 13 days after $B$-band 
maximum, is shown in Fig.~\ref{fig:IR_spec_comp_Ia}, together
with the spectrum of SN~2004aw at +29 days, two spectra of the
BL-SN~1998bw \citep{Patat01} and one of the Type Ia SN~2002bo
\citep{Benetti04}.  The spectrum of SN~1999ex is the only one in
which the lines near $1.04$~$\mu$m and $2.00$~$\mu$m are almost
equally pronounced, suggesting $\mathcal{R}_{\rmn{He}} \approx 1$ if
the $1.04$~$\mu$m absorption is purely He\,{\sc i}, or
$\mathcal{R}_{\rmn{He}} > 1$ if also other elements contribute to it.
In SN~2004aw only the feature at $1.04$~$\mu$m is visible, which could
be interpreted in two ways. The feature being predominantly formed by
He\,{\sc i} would require basic physical conditions near the
photosphere to be substantially different from SN~1999ex in order to
explain the much lower value of $\mathcal{R}_{\rmn{He}} \approx
0$. Models clearly disfavour this option
\citep[see][]{Mazzali98}. Alternatively, the physical conditions are
similar, but the $1.04$~$\mu$m absorption in SN~2004aw is mainly
formed by ions other than He\,{\sc i}.

Apart from the overall narrower features in SN~2004aw, the near-IR
spectrum of this object is quite similar to that of SN~1998bw at a
comparable epoch. This especially holds for the $H$ band.  However, in
the $K$ band SN~2004aw is somewhat reminiscent of SNe~Ia, although all
features are much less pronounced than those in SN~2002bo. Based on near-IR
spectra of the Type Ia SN~1999ee at similar epochs, \citet{Hamuy02}
attributed this group of bumps to blends of iron-group elements. By
contrast, the few near-IR spectra of SNe~Ib/c available so far
seem to be clearly dominated by intermediate-mass elements, even
during the nebular phase \citep{Oliva87,Gerardy_proc}.

\subsection{Ejecta velocities}
\label{Ejecta velocities}

Although already partially discussed in the previous sections, it is
worth taking a more detailed and quantitative look at the expansion
velocities of different chemical species in SN~2004aw. All velocities
have been determined by fitting a Gaussian profile to the absorption
component of the P-Cygni features in the redshift-corrected spectra
and measuring the blueshift of the minimum. The outcome of this
procedure is a rough estimate of the expansion velocities of the
shells where the individual lines predominantly form.  The main caveat
of this method is the fact that most features are actually blends of
several lines, especially when the ejecta velocities are high. In that
case, a shift of the absorption minimum with time could also be
produced by a change in the relative strength of different lines
contributing to the same feature, and hence does not need to be
connected to real velocity evolution due to a receding photosphere.
  
\begin{figure}   
   \centering
   \includegraphics{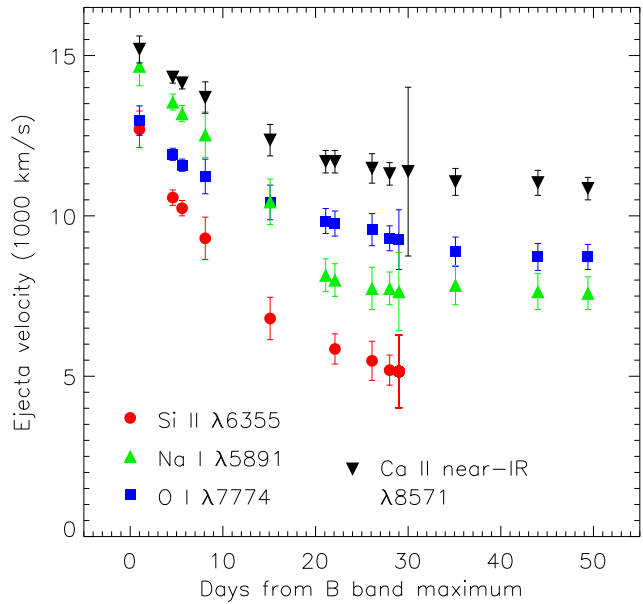}
   \caption{Velocity evolution of different spectral lines of SN~2004aw.}
   \label{fig:velocities}
\end{figure}

Fig.~\ref{fig:velocities} shows the velocity evolution of Si\,{\sc ii}
$\lambda6355$, Na\,{\sc i}\,D, O\,{\sc i} $\lambda7774$, and the
Ca\,{\sc ii} near-IR triplet in SN~2004aw. At all epochs Si\,{\sc ii}
has the lowest and Ca\,{\sc ii} the highest expansion velocities. The
velocities of O\,{\sc i} and Na\,{\sc i}\,D are intermediate and
always similar to each other, but Na\,{\sc i}\,D shows a faster
decline in the early phases until 20 days past maximum. However, for
this line possible contamination with He\,{\sc i} $\lambda5876$ has
to be considered. This may be particularly important at maximum and
soon after when the feature is generally weak, and could partially be
responsible for the apparently different velocity evolution at early
times.

\section{Discussion}
\label{Discussion}

In Section~\ref{Optical light curves} and Fig.~\ref{fig:light curves}
we noted that the delay of maximum light in the red with respect to
bluer bands is one of the remarkable properties of SN~2004aw. The
offset of 8.4 days between the $B$-band and $I$-band maximum
(Table~\ref{phot-quant}) is significantly longer than observed not
only in the ``prototypical" Type Ic SN~1994I (1.88 days;
\citealt{Richmond96}), but also in the Type Ib/c SN~1999ex (6.1 days;
\citealt{Stritzinger02}) and the BL-SNe 1998bw (3.5 days;
\citealt{Galama98}) and 2002ap (5.59 days; \citealt{Foley03}).

\begin{figure*}   
   \centering
   \includegraphics[scale=1.12]{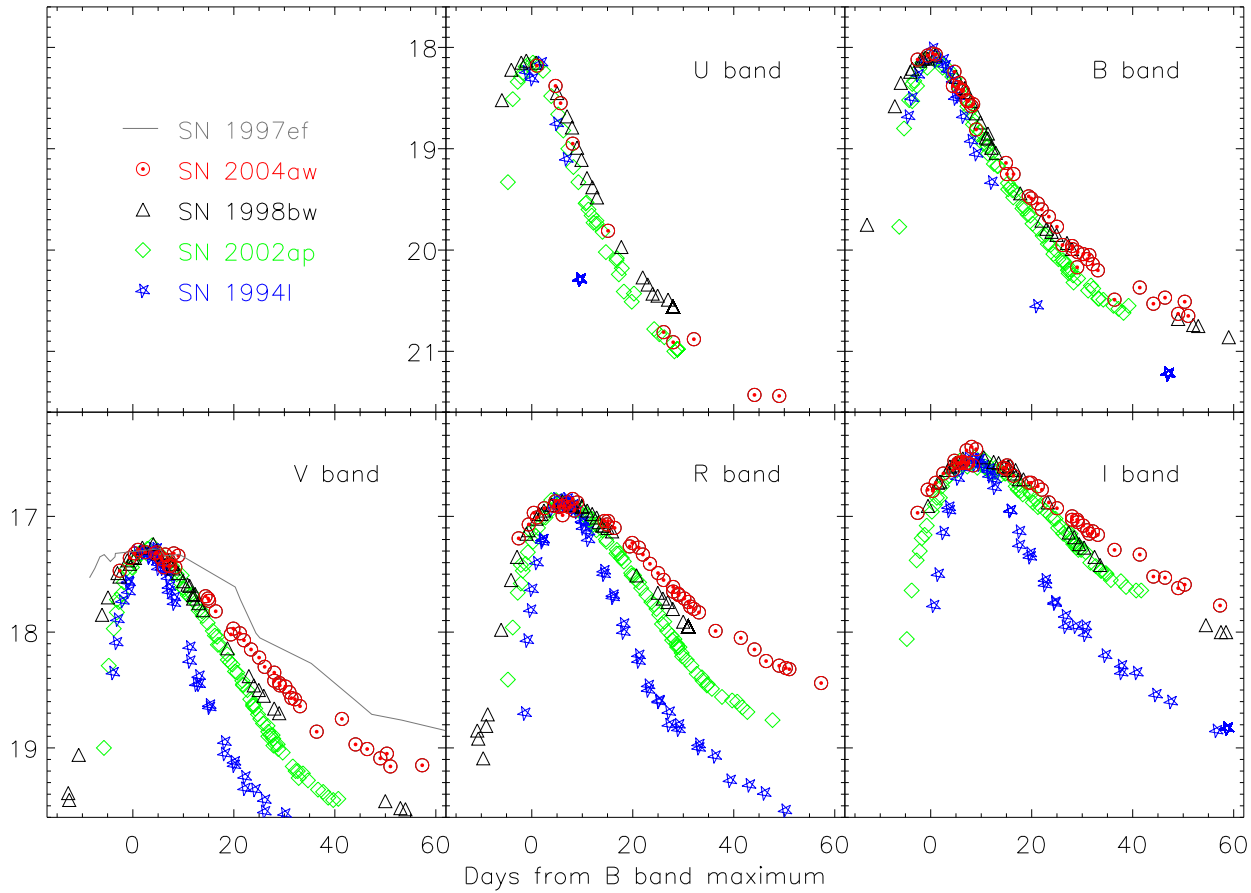}
   \caption{Light curves of SN~2004aw compared with those of SNe
1998bw, 2002ap, 1994I, and 1997ef (see text for references). As in
Fig.~\ref{fig:UBVRI_late}, the latter have been shifted in time and
magnitude to match SN~2004aw at maximum.}
   \label{fig:04aw_98bw_02ap_94I_97ef}
\end{figure*}

Light curves of all SNe~Ic have in common that neither a plateau
with constant luminosity as in SNe~IIP, nor a rebrightening and
secondary maximum in the near-IR bands as in SNe~Ia, is
present. Nevertheless, their exact shape, and in particular the early
decline rate, varies a lot for the different SNe Ic. Some authors have
proposed a bimodal distribution \citep[e.g.][and references
therein]{Clocchiatti,Clocchiatti99}, dividing Type Ic SNe into slow
and fast decliners. Accordingly, SNe~1987M \citep{Filippenko90} and
1994I \citep{Yokoo94,Tsvetkov95,Richmond96} are prominent
representatives of the fast declining group, whereas SNe~1983V
\citep{Clocchiatti97} and 1990B \citep{Clocchiatti01}, the Type Ib/c
SN~1999ex \citep{Stritzinger02}, and the BL-SNe 1998bw
\citep{Galama98,McKenzie99,Patat01,Sollerman02}, 2002ap
\citep{Foley03,Pandey03b,Yoshii03,Tomita06}, 1997ef
\citep{IAUC6778,IAUC6786,IAUC6798,Mazzali04}, and 1997dq
\citep{Mazzali04} belong to the class of slow decliners. However,
Fig.~\ref{fig:04aw_98bw_02ap_94I_97ef}, which compares the light
curves of SN~2004aw with those of other SNe~Ic, suggests instead
that a continuous distribution of decline rates might be a more
appropriate description (see also figs. 4 and 6 of 
\citealt{Richardson06} for a larger sample).

SN~2004aw is one of the slowest decliners of all SNe~Ic, slower than
SN~1998bw and SN~2002ap. Only SN~1997ef and SN~1997dq (the latter is not
included in Fig.~\ref{fig:04aw_98bw_02ap_94I_97ef}) fade even more
slowly, with a decline rate in the radioactive tail that follows the
radioactive decay of $^{56}$Co, indicating little or no $\gamma$-ray
escape \citep{Mazzali04}. On the other hand, the much faster decline
of SN~1994I in all bands is apparent, and quantitatively demonstrated
by its larger $\Delta m_{15}$ ($\Delta m_{15}(B)=2.07$ for
SN~1994I versus $1.09$ for SN~2004aw; see \citet{Richmond96} and
Table~\ref{phot-quant}). Since the width of the light-curve peak is,
assuming the same kinetic energy per unit mass for the two SNe,
closely related to the ejecta mass (with the SN with higher ejecta
mass declining more slowly), the comparison of SNe 2004aw and 1994I
suggests that the progenitor of SN~2004aw was significantly more
massive than that of SN~1994I. We address this topic in more detail in
Section~\ref{Ejecta mass}.

Interestingly, the strongly different decline rates of the light
curves have no direct impact on the colour evolution of the four SNe
included in Fig.~\ref{fig:04aw_98bw_02ap_94I_colors}. The differences
between these objects, emphasised in Section~\ref{Colour
evolution}, appear rather random, and no correlation between the
light-curve decline rates and either the absolute values of the
colours or their time evolution can be established.

In general, diversity seems to be characteristic of SNe
Ic. Correlations between observable parameters -- if they exist -- are
subtle and hence difficult to establish from the small sample of
objects available to date. For example, from the bolometric light
curves in Fig.~\ref{fig:bolom}(a) we find no simple correlation
between the brightness at peak and the decline rate as in Type Ia SNe
\citep[e.g.][]{Phillips99}. Furthermore, we have no evidence that the
expansion velocity of the ejecta and the peak luminosity of a Type Ic
SN are related i.e. BL-SNe might generally be no brighter than
``normal" SNe~Ic. It is true that SN~1998bw not only has extremely
high ejecta velocities, but also is one of the most luminous
core-collapse SNe ever found. In contrast, however, as seen in
Section~\ref{Absolute magnitudes and bolometric light curve},
SNe~2002ap and 1997ef are comparatively faint and at maximum even less
luminous than the normal low-velocity SNe~1994I, 1999ex, and 2004aw.

A comparison of SNe~1994I and 1997ef could lead to the conclusion that
the speed of the spectroscopic evolution of a SN~Ic and the decline
rate of its light curves might be somehow related. SN~1994I shows
rapidly fading light curves and a comparatively fast spectral
evolution (see Fig.~\ref{fig:04aw_98bw_02ap_94I_97ef} and
Section~\ref{Spectroscopic comparison with other SNe Ic}), while
in SN~1997ef a very slow light-curve decline comes along with an
unusually persistent continuum in the spectra
\citep{Mazzali04}. However, hopes of a useful correlation are
diminished by the Type Ic SN~1990B \citep{Clocchiatti01}, which
exhibited slowly declining light curves yet a fast spectroscopic
evolution.\\
 
In Section~\ref{Spectroscopic comparison with other SNe Ic} we
showed that from a spectroscopic point of view SN~2004aw shares
properties with both the group of so-called ``normal" SNe~Ic (e.g.,
SNe 1983V, 1990B, and of course 1994I) and the moderate BL-SNe 1997ef,
2002ap and especially 2003jd. This intermediate position -- together
with the wide variety of characteristics among BL-SNe themselves --
gives rise to the question of whether BL-SNe actually form a distinct
subclass, or if it is more appropriate to think of a continuity of
properties from ordinary low-velocity SNe~Ic to the most powerful
events like SN~1998bw.

An idea to explain the origin of the observed diversity has been to
assume that broad-lined SNe are highly aspherical, jet-driven events
(some of them connected to a GRB), while ordinary SNe~Ic represent
more spherical explosions \citep{Mazzali05b}. In order to verify this
picture, the examination of nebular spectra is of particular interest,
as they provide an insight into the innermost layers of the ejecta and
the geometry of the explosion. \citet{Mazzali05b} discussed what
GRB-related SNe may look like when they are observed from directions
almost perpendicular to the jet, so that the burst itself is not
visible. Their prediction was that SNe seen along the jet axis have a
narrow, sharp, single-peaked [O\,{\sc i}] $\lambda\lambda6300,6364$
and strong and comparatively broad [Fe\,{\sc ii}] emission features,
while if viewed off-axis the [Fe\,{\sc ii}] feature should be narrower
and weaker, and the [O\,{\sc i}] double-peaked. Finally, spherical
explosions without a jet should produce a nebular [O\,{\sc i}] line
with a flat or rounded top. 

As can be seen from Fig.~\ref{fig:comp_plus236}, SN~1998bw fits well
into the scenario of a GRB-SN viewed fairly along the jet axis,
consistent with the detection of the spatially and temporally
coincident GRB 980425. SN~2003jd exhibits a double-peaked [O\,{\sc i}]
line, indicating that it might be a GRB-SN viewed well off-axis
(\citealt{Mazzali05b}, but see also \citealt{Soderberg06}). With its
rounded tops of the emission peaks, SN~1994I might represent a rather
spherical explosion.  SN~2002ap and SN~2004aw show the same sharp
[O\,{\sc i}] line core as SN~1998bw, but in both cases no GRB was
detected.  Following the arguments of \citet{Maeda02}, \citet{Tomita06}
and Mazzali et al. (in preparation), the observed line profiles
require both explosions to be strongly asymmetric, but maybe the jet
was too strongly contaminated by baryons to trigger a GRB
\citep{Tominaga04}. However, this still does not explain the
completely different kinetic energy per unit mass in the outer ejecta
of SN~2004aw and SN~2002ap, as indicated by the significantly
different line widths at early phases.

\subsection{Search for an associated GRB}
\label{associated GRB}

In order to probe the possible association of SN~2004aw with a GRB the 
data of the Interplanetary Network (IPN\footnote{\texttt{http://www.ssl.berkeley.edu/ipn3/}}) 
of high-energy satellites (see, e.g., \citealt{Hurley99}) were examined 
in the period UT March 2 to 17, i.e. from about 7 to 22 days prior to 
maximum light in $B$. At that time the IPN involved six spacecraft 
(Konus-Wind, HETE, Mars Odyssey, RHESSI, RXTE, and INTEGRAL), securing 
nearly isotropic coverage and almost 100\% duty cycle. 

A total of seven confirmed GRBs (i.e., events observed by at least two 
experiments, and having GRB characteristic), plus three unconfirmed 
events (observed by a single instrument, but likely being genuine GRBs 
and not particle-induced events) were identified (K. Hurley, private 
communication). For six out of these ten candidates the localisation 
through triangulation was sufficiently precise to exclude a spatial 
association with SN~2004aw. For the remainder (two confirmed and two 
unconfirmed bursts) there was at best a very coarse localisation 
(excluding only 30 to 85\% of the sky). Within this limited accuracy 
the four events are all consistent with SN~2004aw, although the 
probability that this is by chance is very large.

Hence, the result of the search remains rather inconclusive, as an 
association of SN~2004aw with a GRB can be neither firmly established 
nor definitely excluded. Assuming that none of the unconfirmed GRBs
was associated with SN~2004aw, a rough upper limit of 
$10^{-6}$\,erg\,\,cm$^{-2}$ can be set on the fluence of a possible 
accompanying GRB in the energy range 25 -- 150 keV. However, depending 
on the spectrum and time history of the burst, the fluence threshold 
could vary by an approximate factor of 5 in either direction.

\subsection{Mass of the ejecta}
\label{Ejecta mass}

\begin{table*}
\caption{Comparison of parameters of SNe~Ic.$^a$} 
\label{parameters}
\begin{center}
\begin{footnotesize}
\begin{tabular}{lccccl}
\hline
SN          & M$_V$ (mag) & $E_{kin} / 10^{51}$erg & $M_{ej}/M_\odot$ & $M_{N\!i}/M_\odot$ & References \\
\hline
1994I       & $-17.62$    & 1                      & 0.9              & 0.07               & \citealt{Richmond96,Nomoto94} \\
2004aw      & $-18.02$    & 3.5--9.0               & 3.5--8.0         & 0.25--0.35         & this paper \\
2002ap      & $-17.35$    & 4                      & 3                & 0.08               & \citealt{Foley03,Mazzali02,Tomita06} \\
1997ef      & $-17.14$    & 19                     & 9.5              & 0.16               & \citealt{Mazzali00,Mazzali04} \\
1998bw      & $-19.13$    & 30                     & 10               & 0.70               & \citealt{Galama98,Nakamura00} \\
\hline
\end{tabular}
\\[1.6ex]
$^a$ For all SNe except SN~2004aw, the values for kinetic energy, ejecta
mass, and nickel mass have been inferred from light-curve and spectral
models.
\end{footnotesize}
\end{center} 
\end{table*}
In order to roughly estimate the total mass of the ejecta of
SN~2004aw, we make use of an analytical description of the peak of the
light curve developed by \citet{Arnett82}. The model assumes spherical
symmetry, a homologous expansion, no mixing of $^{56}$Ni, a constant
optical opacity $\kappa_{opt}$, radiation-pressure dominance, and the
applicability of the diffusion approximation for photons, which
restricts it to early phases when the density is sufficiently high to
make the ejecta optically thick. In this model the time evolution of
the SN luminosity is given by
\begin{equation}
L(t) = \varepsilon_{N\!i}\,M_{N\!i}\ e^{-x^2}\int_0^x 2z\,e^{-2zy+z^2}dz,
\label{L}
\end{equation}
where $x\equiv\frac{t}{\tau_m}$, $y\equiv\frac{\tau_m}{2\tau_{N\!i}}$,
and $\varepsilon_{N\!i}=\frac{Q_{N\!i}}{m_{N\!i}\tau_{N\!i}}$.\\ Here
$Q_{N\!i}$ is the energy release per $^{56}$Ni decay (1.73 MeV, not
taking into account the contribution of neutrinos, for which the
environment is optically thin), and $\tau_{N\!i}$ is the $e$\,-folding
time of the $^{56}$Ni decay (8.77 days). The effective diffusion time
$\tau_m$ determines the width of the peak of the bolometric light
curve and is given by
\begin{equation}
\tau_m \, = \, \left(\frac{2}{\beta c}\,\frac{\kappa_{opt}M_{ej}}{v_{sc}}\right)^{\!1/2} \propto\ \kappa_{opt}^{1/2}\,M_{ej}^{3/4}\,E_{kin}^{-1/4},
\label{tau}
\end{equation}
where $v_{sc}$ is the velocity scale of the homologous expansion and
$\beta$ is an integration constant. Given a certain light-curve shape
(i.e., a certain value of $\tau_m$), Eq.~\ref{tau} allows for a family
of photometrically degenerate models with different combinations of
$M_{ej}$ and $E_{kin}$. By specifying $E_{kin}$ or, alternatively, the
velocity of the ejecta, this degeneracy can be broken, and an estimate
of the total ejecta mass becomes feasible. For our mass estimate we
make use of Eq.~\ref{tau} and compare SN~2004aw to SN~1994I, for which
the relevant parameters are well known from detailed modelling
\citep{Nomoto94}.  We do not consider any differences in the mean
optical opacities $\kappa_{opt}$.

The comparison to SN~1994I starts from the observation that the ejecta
velocities of these two objects are quite similar
($v_{sc,\rmn{04aw}}\approx v_{sc,\rmn{94I}}$), but that the widths of
the bolometric light curves are significantly different. Taking
$\tau_m$ as the time from peak of the bolometric light curve to the
moment when the luminosity is equal to $1/e$ times the peak luminosity
(which is equivalent to a decline of 1.1 mag), we find that
$\tau_{m,\rmn{04aw}}/\tau_{m,\rmn{94I}}\approx 2.0$--$3.0$ (the large
uncertainty arises from the unknown effect of the bolometric
corrections when considering also the near-IR bands). Assuming
$v_{sc,\rmn{04aw}}\approx v_{sc,\rmn{94I}}$ we can solve Eq.~\ref{tau}
for $M_{ej,\rmn{04aw}}$ and obtain
\begin{equation}
M_{ej,\rmn{04aw}} \approx M_{ej,\rmn{94I}}
\left(\frac{\tau_{m,\rmn{04aw}}}{\tau_{m,\rmn{94I}}}\right)^{\!2}
\approx 4.0-9.0\times M_{ej,\rmn{94I}}.
\end{equation}
With $M_{ej,\rmn{94I}}=0.9\,M_\odot$ \citep{Nomoto94} the range of possible
ejecta masses for SN~2004aw turns out to be $3.5$--$8.0\,M_\odot$. The
corresponding kinetic energy, computed via $v_{sc,\rmn{04aw}}\approx
v_{sc,\rmn{94I}}$ with $E_{kin,\rmn{94I}}\approx
1.0\,\times\,10^{51}$erg \citep{Nomoto94}, would range from 3.5 to
9.0$\,\times\,10^{51}$erg. If instead $M_{ej,\rmn{94I}}=1.2\,M_\odot$ 
as suggested by \citet{Sauer06} was adopted, both the inferred ejecta mass 
and the kinetic energy of SN~2004aw would increase by about 30\%.

Hence, this comparison suggests an ejecta mass for SN~2004aw of at
least 3.5 $M_\odot$. This value is substantially larger than that found for
SN~1994I, and even larger than that for the BL-SN~2002ap (for a comparison
of explosion properties see Table~\ref{parameters}). At the same time
the kinetic energy is less enhanced with respect to SN~2002ap,
consistent with the ejecta velocities in SN~2004aw being significantly lower 
than those in SN~2002ap.\\

Eq.~\ref{L} can also be utilised to estimate the mass of $^{56}$Ni
synthesised in the explosion. For this purpose the formula was
resolved for $M_{N\!i}$ and evaluated at the time of peak
brightness. The effective diffusion time $\tau_m$ was determined as
described above, and for $L$ the value from our quasi-bolometric 
($U$-through-$I$) light curve was adopted. To account for the necessary
corrections to the true bolometric luminosity and to ensure that the
zero-point of the Arnett (1982) relation is properly fixed, we repeated
these steps for SNe~1994I, 1998bw, and 2002ap, and introduced an
additional factor of 1.1 in Eq.~\ref{L} that enabled us to better reproduce
the Ni masses reported in Table~\ref{parameters} for these
objects. What we finally find for SN~2004aw is a $^{56}$Ni mass of
about $0.30\pm0.05\,M_\odot$, well above the average for core-collapse
supernovae.\\

Given the comparatively high ejecta mass and Ni production in
SN~2004aw, it is interesting to discuss which physical parameters of
the progenitor star are crucial for accelerating the ejecta to the
high velocities observed in BL-SNe. If we assume that the ejecta mass
is correlated with the total mass of the progenitor star just before
core collapse (which is surely true if the compact remnant is a
neutron star with a strict upper mass limit), then the progenitor of
SN~2004aw was probably more massive than that of SN~2002ap. But even
if the possibility of black hole formation in at least one of the two
objects is taken into account, the probability that fall-back of
matter is sufficiently strong to invert the progenitor mass versus
ejecta mass relation is low. This almost certainly rules out the
progenitor mass as the only criterion for producing broad-lined SNe.
Since BL-SNe are supposed to be highly aspherical jet-like explosions,
it is plausible that the angular momentum of the progenitor star (or
the progenitor system in case of a binary) will also play an important
role \citep[see e.g.][and references therein]{Woosley06}.\\

\section{Summary and Outlook}
\label{Summary}

We have presented a comprehensive set of optical and near-IR data of
the Type Ic SN~2004aw. The light curves of this object, both in the
optical and the IR, show a peak broader than that observed in most SNe~Ic
so far. Also, the decline during the radioactive tail is rather slow,
suggesting an enhanced trapping of $\gamma$-rays with respect to the
majority of SNe~Ic. Both findings require a relatively large ejecta
mass, which in the previous section was determined to be 3.5--8.0
$M_\odot$. With absolute peak magnitudes of about $-18$ in all optical
bands, SN~2004aw seems to be a bit brighter than the average of all
SNe~Ic, but still more than 1 mag underluminous with respect to the
very bright BL-SN~1998bw.  The optical spectra show the typical lines
found in most SNe~Ic. Line blending is similar to that seen in spectra
of SN~1994I and much weaker than in the BL-SNe, owing to normal ejecta
velocities. The spectroscopic evolution is rather slow, making the
moderate BL-SNe 1997ef and 2003jd a better match than SN~1994I a few
weeks past maximum. The nebular [O\,{\sc i}] $\lambda\lambda6300,6364$
line shows a sharp, narrow core, as would be expected for a highly
aspherical explosion. Unfortunately, no final conclusion can be drawn 
concerning a possible association with a GRB, because the localisation 
of some burst candidates is not sufficiently precise. Also the detection 
of He remains controversial, but we can at least say that in SN~2004aw 
the feature observed at $1.04$~$\mu$m cannot be helium alone.

With the physical parameters computed in the previous section,
SN~2004aw proves to be a fairly massive and energetic representative
of the SN~Ic class. About 0.3 $M_\odot$ of synthesised $^{56}$Ni is a
large amount for a core-collapse SN, and both ejecta mass and total
kinetic energy are a factor of 4 to 9 larger than in the prototypical
Type Ic SN~1994I. This makes SN~2004aw one of the most energetic
normal-velocity SNe~Ic ever observed, lying in a parameter region so
far almost exclusively populated by some BL-SNe.

These properties also confirm that the various definitions of
``hypernova" mentioned in Section~\ref{Introduction} are not
equivalent. We do believe that SN~2004aw was a highly aspherical
explosion, but the ejecta velocities were normal and the total kinetic
energy was fairly high, though still below $10^{52}$ erg. However, the
latter also holds for SN~2002ap (see Table~\ref{parameters}), which is
by some authors referred to as hypernova owing to its high ejecta
velocities and SN~1998bw-like spectra near maximum light \citep[see
e.g.][]{Mazzali02,Pandey03a}.

What we believe to see in both normal and broad-lined SNe~Ic is a
continuous range of the most relevant physical parameters such as
nickel mass, ejecta mass, and kinetic energy. A wide variety of
combinations of these parameters is actually realised in nature,
giving rise to the tremendous diversity in the Type Ic SN subclass.

\section*{Acknowledgments}

This work has been supported by the European Union's Human Potential
Programme ``The Physics of Type Ia Supernovae," under contract
HPRN-CT-2002-00303. A.V.F.'s group at UC Berkeley is supported by
National Science Foundation (NSF) grant AST-0307894, and by a generous
gift from the TABASGO Foundation. A.G. and V.S. would like to thank
the G\"oran Gustafsson Foundation for financial support.

This paper is based on observations collected at the 2.2\,m and the
3.5\,m Telescopes of the Centro Astron\'omico Hispano Alem\'an (Calar
Alto, Spain), the Asiago 1.82\,m Telescope (INAF Observatories, Italy),
the Telescopio Nazionale Galileo, the Nordic Optical Telescope and the
William Herschel Telescope (La Palma, Spain), the ESO Very Large
Telescope (Cerro Paranal, Chile), the Katzman Automatic Imaging
Telescope and the Shane 3\,m reflector (Lick Observatory, California,
USA), the 10\,m Keck-I Telescope and the United Kingdom Infrared
Telescope (Mauna Kea, Hawaii, USA), the 2.3\,m Telescope (Siding Spring
Observatory, Australia), and the Wendelstein 0.8\,m Telescope (Bavaria,
Germany). The W. M. Keck Observatory is operated as a scientific
partnership among the California Institute of Technology, the
University of California, and NASA; it was made possible by the
generous financial support of the W. M. Keck Foundation. KAIT owes its
existence to donations from Sun Microsystems, Inc., the
Hewlett-Packard Company, AutoScope Corporation, Lick Observatory, the
NSF, the University of California, the Sylvia \& Jim Katzman
Foundation, and the TABASGO Foundation. The United Kingdom Infrared
Telescope is operated by the Joint Astronomy Centre on behalf of the
U.K. Particle Physics and Astronomy Research Council. The UKIRT data
reported here were obtained as part of the UKIRT Service
Programme. The Asiago 1.82\,m Telescope is operated by the INAF --
Osservatorio Astronomico di Padova.

Our thanks go to the support astronomers at the Telescopio Nazionale
Galileo, the 2.2\,m and 3.5\,m Telescopes in Calar Alto, the ESO Very
Large Telescope, and the United Kingdom Infrared Telescope for
performing the follow-up observations of SN~2004aw. We are grateful to
Roberto Nesci for his collaboration during the ToO observations of
SN~2004aw with the Asiago 1.82\,m Telescope, and to Mohan
Ganeshalingam, David Pooley, and Diane S. Wong for assistance with
some of the observations at Lick and Keck. We also thank Kari Nilsson
and G\"oran \"Ostlin for giving up part of their for giving up part of
their time at the Nordic Optical Telescope, and Thomas Augusteijn,
Tapio Pursimo, Erik Zackrisson, and Daniel Arnberg for carrying out
the observations. S.T. is indebted to Heinz Barwig for arranging the
observations at the Wendelstein Observatory.
Moreover, we are grateful to K. Hurley for the Interplanetary Network 
information, and to the scientists of the Konus-Wind, HETE, Mars 
Odyssey (HEND and GRS), RHESSI, RXTE ASM, and INTEGRAL (SPI-ACS) 
experiments for contributing their data to the network.

This research made use of the NASA/IPAC Extragalactic Database (NED)
which is operated by the Jet Propulsion Laboratory, California
Institute of Technology, under contract with the National Aeronautics
and Space Administration, and the Lyon-Meudon Extragalactic Database
(LEDA), supplied by the LEDA team at the Centre de Recherche
Astronomique de Lyon, Observatoire de Lyon. This publication also
used data products from the Two Micron All Sky Survey (2MASS), which
is a joint project of the University of Massachusetts and the Infrared
Processing and Analysis Center/California Institute of Technology,
funded by the National Aeronautics and Space Administration and the
National Science Foundation.

Last but not least we want to thank the referee of this paper for his 
constructive comments.

\addcontentsline{toc}{chapter}{Bibliography}
\markboth{Bibliography}{Bibliography}
\bibliographystyle{mn2e}


\label{lastpage}

\end{document}